\documentclass[journal]{IEEEtran} 
\IEEEoverridecommandlockouts

\newcommand{\figwid}{90mm}
\usepackage{graphicx,psfrag}
\usepackage{amsbsy,amsmath,amsfonts,amssymb,mathtools,bbm}
\usepackage{cite}
\usepackage{hyperref,url}
\usepackage{algorithm,algpseudocode}
\usepackage{algcompatible}
\usepackage{multicol,booktabs}
\usepackage{diagbox}[2011/11/22]
\usepackage{tikz}
\usepackage{relsize}
\usepgflibrary{patterns}
\usetikzlibrary{arrows,positioning,fit,shapes,calc,decorations.pathreplacing,matrix,patterns}
\tikzset{block/.style={draw,thick,text width=1.6cm,minimum height=0.8cm,align=center},
         line/.style={-latex}
}

\newcommand{\putTable}[3]{\begin{table}[tp]
		\centering 			        
		\caption{#2}
		\vspace{-4mm}
		#3
		\label{tab:#1}
\end{table} }

\newcommand{\re}{^{\text{re}}}
\newcommand{\im}{^{\text{im}}}

\renewcommand{\tilde}{\widetilde}
\renewcommand{\hat}{\widehat}
\renewcommand{\bar}{\overline}
\newcommand*\dif{\mathop{}\!\mathrm{d}}

\newcommand{\0}{\boldsymbol{0}}
\newcommand{\1}{\boldsymbol{1}}
\newcommand{\bb}{\boldsymbol{b}}

\newcommand{\cc}{\boldsymbol{c}}

\newcommand{\e}{\boldsymbol{e}}
\newcommand{\f}{\boldsymbol{f}}
\newcommand{\g}{\boldsymbol{g}}
\newcommand{\h}{\boldsymbol{h}}

\newcommand{\p}{\boldsymbol{p}}
\newcommand{\q}{\boldsymbol{q}}
\newcommand{\qhat}{\hat{\boldsymbol{q}}}
\newcommand{\rhat}{\hat{\boldsymbol{r}}}
\newcommand{\s}{\boldsymbol{s}}
\newcommand{\uu}{\boldsymbol{u}}
\newcommand{\vv}{\boldsymbol{v}}
\newcommand{\w}{\boldsymbol{w}}
\newcommand{\x}{\boldsymbol{x}}
\newcommand{\y}{\boldsymbol{y}}
\newcommand{\z}{\boldsymbol{z}}

\newcommand{\A}{\boldsymbol{A}}
\newcommand{\F}{\boldsymbol{F}}
\newcommand{\G}{\boldsymbol{G}}
\newcommand{\Hmat}{\boldsymbol{H}}
\newcommand{\I}{\boldsymbol{I}}
\newcommand{\J}{\boldsymbol{J}}
\newcommand{\Pmat}{\boldsymbol{P}}
\newcommand{\Q}{\boldsymbol{Q}}
\newcommand{\Smat}{\boldsymbol{S}}
\newcommand{\U}{\boldsymbol{U}}

\newcommand{\W}{\boldsymbol{W}}
\newcommand{\X}{\boldsymbol{X}}

\newcommand{\Z}{\boldsymbol{Z}}

\newcommand{\bnu}{\boldsymbol{\nu}}
\newcommand{\bmu}{\boldsymbol{\mu}}
\newcommand{\blam}{\boldsymbol{\lambda}}

\newcommand{\xP}{\boldsymbol{x}_\textrm{P}}
\newcommand{\xD}{\boldsymbol{x}_\textrm{D}}
\newcommand{\xG}{\boldsymbol{x}_\textrm{G}}
\newcommand{\xCP}{\boldsymbol{x}_\textrm{C}}

\newcommand{\nux}{\nu^{\text{x}}}
\newcommand{\nuh}{\nu^{\text{h}}}
\newcommand{\nup}{\nu^{\text{p}}}
\newcommand{\barnup}{\bar{\nu}^{\text{p}}}
\newcommand{\nuq}{\nu^{\text{q}}}
\newcommand{\nur}{\nu^{\text{r}}}
\newcommand{\nuz}{\nu^{\text{z}}}
\newcommand{\nus}{\nu^{\text{s}}}

\newcommand{\Nd}{N_\textrm{D}}

\newcommand{\Ng}{N_\textrm{G}}
\newcommand{\Nb}{N_\textrm{b}}
\newcommand{\Ncp}{N_\textrm{C}}
\newcommand{\Kd}{K_\textrm{D}}
\newcommand{\Kp}{K_\textrm{P}}

\newcommand{\Gy}{\G_{\textrm{\bf y}}}
\newcommand{\Gx}{\G_{\textrm{\bf x}}}
\newcommand{\Gw}{\G_{\textrm{\bf w}}}

\newcommand{\giv}{\,|\,}

\DeclareMathOperator{\tr}{tr}
\newcommand{\herm}{^{\textsf{H}}}
\newcommand{\tran}{^{\textsf{T}}}
\newcommand{\Complex}{\mathbb{C}}
\newcommand{\Real}{\mathbb{R}}

\newcommand{\Ex}{\mathbb{E}}
\newcommand{\diag}{\textrm{diag}}
\newcommand{\Diag}{\textrm{Diag}}
\newcommand{\vvec}{\textrm{vec}}

\newcommand{\RRe}{\text{Re}}
\newcommand{\IIm}{\text{Im}}

\newcommand{\CN}{\mathcal{CN}}

\newcommand{\rvp}{\textsf{p}}
\newcommand{\rvy}{\textsf{y}}
\newcommand{\rvyy}{\textsf{\textbf{y}}}
\newcommand{\rvz}{\textsf{z}}
\newcommand{\rvzz}{\textsf{\textbf{z}}}
\newcommand{\rvh}{\textsf{h}}
\newcommand{\rvhh}{\textsf{\textbf{h}}}

\newcommand{\rvc}{\textsf{c}}
\newcommand{\rvcc}{\textsf{\textbf{c}}}
\newcommand{\rvx}{\textsf{x}}
\newcommand{\rvxx}{\textsf{\textbf{x}}}
\newcommand{\rvq}{\textsf{q}}
\newcommand{\rvr}{\textsf{r}}
\newcommand{\rvw}{\textsf{w}}

\newcommand{\sigtil}{\sigma_{\tilde{w}}^2}

\newcommand{\Figref}[1]{Figure~\ref{fig:#1}}
\newcommand{\figref}[1]{Fig.~\ref{fig:#1}}
\newcommand{\tabref}[1]{Table~\ref{tab:#1}}

\newcommand{\secref}[1]{Sec.~\ref{sec:#1}}
\DeclareMathOperator{\var}{var}
\newcommand{\defn}{\triangleq}

\newcommand\minus{%
  \setbox0=\hbox{-}%
  \vcenter{%
    \hrule width\wd0 height \the\fontdimen8\textfont3%
  }%
}

\begin{document}

\title{Joint Channel-Estimation/Decoding with Frequency-Selective Channels and Few-Bit ADCs}

\author{Peng Sun, Zhongyong Wang, 
        Robert W. Heath, Jr., \emph{Fellow, IEEE}, and 
        Philip Schniter,\IEEEauthorrefmark{1} \emph{Fellow, IEEE}%
        \thanks{P. Sun (sun.1771@osu.edu) and Z. Wang (iezywang@zzu.edu.cn) are with the 
                School of Information Engineering, Zhengzhou University, Zhengzhou 450001, China.
                This work was performed while P. Sun was visiting the Department of ECE at 
                The Ohio State University, Columbus, OH.
                }
        \thanks {R. Heath (rheath@utexas.edu) is with the 
                Wireless Networking and Communications Group, The University of Texas at Austin, Austin, TX 78712, USA.}
        \thanks {P. Schniter (schniter.1@osu.edu) is with the 
                Department of ECE at The Ohio State University, Columbus, OH 43210, USA. 
                Please direct all correspondence to Prof.\ Philip Schniter,
                Dept. ECE, 2015 Neil Ave., Columbus, OH 43210.
                phone 614.247.6488, fax 614.292.7596.}%
        \thanks{This work was supported in part by 
                the National Science Foundation under grants CCF-1527079 and CCF-1527162,
                and the National Natural Science Foundation of China under grant NSFC-61571402.}%
        \thanks{Portions of this work were presented at 2017 Asilomar Conference on Signals, Systems, and Computers.}%
        }

\maketitle
\IEEEpeerreviewmaketitle

\begin{abstract}
We propose a fast and near-optimal approach to joint channel-estimation, equalization, and decoding of coded single-carrier (SC) transmissions over frequency-selective channels with few-bit analog-to-digital converters (ADCs).
Our approach leverages parametric bilinear generalized approximate message passing (PBiGAMP) to reduce the implementation complexity of joint channel estimation and (soft) symbol decoding to that of a few fast Fourier transforms (FFTs).
Furthermore, it learns and exploits sparsity in the channel impulse response.
Our work is motivated by millimeter-wave systems with bandwidths on the order of Gsamples/sec, where few-bit ADCs, SC transmissions, and fast processing all lead to significant reductions in power consumption and implementation cost.
We numerically demonstrate our approach using signals and channels generated according to the IEEE 802.11ad wireless local area network (LAN) standard, in the case that the receiver uses analog beamforming and a single ADC.
\end{abstract}


\section{Introduction} \label{sec:intro}

The trend towards ever-wider-bandwidths in communications systems results in major implementational challenges.
This trend is evident in millimeter-wave (mmWave) systems, which exploit large chunks of bandwidth at carrier frequencies of 30~GHz and above \cite{Heath:JSTSP:16}.
For example, the IEEE~802.11ad standard \cite{802.11ad} specifies channels of bandwidth 1.76~GHz centered near 60~GHz.
Future 5G cellular systems are also likely to incorporate mmWave technology \cite{Rangan:Proc:14,Rappaport:IA:13}.

A main challenge in wideband systems comes from the analog-to-digital converters (ADCs) used at the receiver.
At bandwidths above 1~Gs/sec, ADC power consumption grows approximately quadratically with bandwidth \cite{Murmann:FTFC:13,Murmann:18}.
Meanwhile, ADC power consumption grows exponentially in the number of bits used in conversion.
At GHz bandwidths, many-bit (e.g., 10~bit) ADCs may consume several watts of power, which is impractical for handheld mobile devices. 
For this reason, there has been a growing interest in few-bit (i.e., 1-4~bit) ADCs for communications receivers (e.g.,
\cite{Mezghani:WSA:12, Mo:ASIL:14, Jacobsson:TWC:17, Mezghani:TSP:18, Zhou:WCNC:17, 
      Mezghani:WSA:07, Mezghani:ICASSP:08, Mezghani:ISIT:10, SCWang:TWC:15, Xiong:WCNC:17,
      Wen:TSP:16, Steiner:WSA:16,
      Dabeer:ICC:10, Zeitler:TSP:12,
      Mezghani:WSA:10,
      Li:TSP:17,
      Lok:ISIT:98,
      Mo:TSP:18,
      SCWang:ICC:14a}).

Wide bandwidth also results in challenges at the transmitter. 
In particular, wide-bandwidth linear amplifiers are expensive in terms of power consumption and cost \cite{Falconer:CM:02}.
For this reason, it is beneficial to transmit signals with low peak-to-average power ratio (PAPR), which allow power-amplifier linearity requirements to be relaxed.
The desire for low PAPR suggests single-carrier (SC) transmission, as opposed to multi-carrier transmission such as orthogonal frequency division multiplexing (OFDM) \cite{Bingham:CM:90}. 
Because wide bandwidth receivers may need to decode billions of bits per second, it is important that the SC transmission is amenable to computationally efficient channel-equalization, e.g., via fast Fourier transform (FFT) processing \cite{Falconer:CM:02}. 

Although wide bandwidth brings many challenges, there is a silver lining: 
the measured channel responses are relatively sparse in the angle and delay domains, in both indoor \cite{Maltsev:JSAC:09} and outdoor \cite{Rappaport:TAP:13,Akdeniz:JSAC:14} settings.
With sparse channels, the fundamental performance of a communications link can be significantly improved (e.g., \cite{Kannu:TIT:11,Schniter:ASIL:14}). 

We now review relevant existing work on few-bit-ADC receiver design.
For flat-fading multiple-input/multiple-output (MIMO) channels, 
channel estimation
(e.g., \cite{Mezghani:WSA:12, Mo:ASIL:14, Jacobsson:TWC:17, Mezghani:TSP:18, Zhou:WCNC:17}),
symbol detection
(e.g., \cite{Mezghani:WSA:07, Mezghani:ICASSP:08, Mezghani:ISIT:10, SCWang:TWC:15, Xiong:WCNC:17}),
and joint channel estimation and symbol detection
(e.g., \cite{Wen:TSP:16, Steiner:WSA:16})
have been considered.
However, wideband channels are frequency selective in practice. 
 
For frequency-selective channels, channel
estimation has been considered in \cite{Dabeer:ICC:10, Zeitler:TSP:12} using comb-type pilots that allow the channel to be treated as effectively flat-fading, but these approaches perform poorly under PAPR limits.
Channel estimation for 2-tap channels was considered in \cite{Mezghani:WSA:10}, but realistic wideband channels have many more taps.
An approach for longer channels was recently proposed in \cite{Li:TSP:17}, but it applies only to OFDM.
An iterative expectation-maximization (EM)-like channel estimation scheme for SC transmissions was proposed in \cite{Lok:ISIT:98}, but it is computationally expensive and does not leverage sparsity.
More recently, pilot-aided sparsity-exploiting channel-estimation schemes were proposed in \cite{Mo:TSP:18},
and a known-channel symbol-detection scheme was proposed in \cite{SCWang:ICC:14a}.
Both \cite{Mo:TSP:18} and \cite{SCWang:ICC:14a} are made computationally efficient by the use of 
generalized approximate message passing (GAMP) \cite{Rangan:ISIT:11} and FFT processing.
But, as we will show, significantly improved performance can be obtained through \emph{joint} channel estimation, symbol detection, and bit decoding.
A joint channel-estimation/decoding approach was proposed in \cite{Cao:ICNC:17}, but it does not leverage sparsity and requires OFDM.

In this paper, we propose a computationally efficient approach to joint channel-estimation, equalization, and decoding of single-carrier transmissions over frequency-selective channels with few-bit ADCs.
Our approach is an instance of turbo-equalization \cite{Douillard:ETT:95,Koetter:SPM:04},
which iterates soft equalization (and, in our case, joint channel estimation) with soft decoding.
For joint channel estimation and equalization, we use the recently proposed Parametric Bilinear GAMP (PBiGAMP) framework \cite{Parker:JSTSP:16}, which---when specialized to our application---consumes only a few FFTs per equalizer iteration and demands relatively few equalizer iterations.
We then mate PBiGAMP to the soft decoder using the turbo-AMP framework from \cite{Schniter:CISS:10}.
To exploit the channel's (approximate) sparsity, we use a Gaussian mixture model (GMM), as in \cite{Schniter:JSTSP:11}, and learn the GMM parameters via the EM algorithm, building on \cite{Vila:TSP:13}.
Portions of this work were published in \cite{Sun:ASIL:17}.
Relative to \cite{Sun:ASIL:17}, this paper includes 
detailed derivations and explanations, 
a refined channel-estimation scheme, 
and additional numerical experiments.

In this work, we assume the use of analog beamforming, and thus a single (few-bit) ADC, at the receiver. 
Our approach can be contrasted with digital (e.g., \cite{Mo:TSP:18}) or hybrid (e.g., \cite{Alkhateeb:JSTSP:14}) beamforming, which requires the use of multiple ADCs.
It is possible that, for large arrays, with our architecture, the power consumption of the analog beamforming becomes more significant than that of the ADCs; 
The exact calculation is architecture-specific (see, e.g., \cite{Yan:18}) and we leave an investigation of these issues to future work.
Extensions of our approach to digital beamforming systems and to hybrid analog/digital systems are worthwhile, but outside the scope of this work.
To evaluate our receiver design, we consider a system that complies with the IEEE~802.11ad 60~GHz mmWave standard \cite{802.11ad}, which supports analog beamforming.
Our numerical results for the IEEE 802.11ad ``conference room'' channel \cite{Maltsev:Tech:10} (under perfect synchronization) show only a 3dB SNR gap at a BER of $10^{-2}$ for a 2-bit ADC compared to infinite bit resolution also using joint decoding. 
Further, we show how embracing the nonlinearity of the quantization helps to avoid a substantial SNR gap that arises when pilot-only channel estimation is used or when Bussgang linearization is used with very-few-bit ADCs at high SNR.

The paper is organized as follows. 
In \secref{sysmodel}, we present our models for
SC block transmission, channel propagation, and few-bit reception,
as well the GMM-based channel model that we use with PBiGAMP.
In \secref{JCEDpbigamp}, after a brief introduction to belief propagation and PBiGAMP, we propose our soft joint channel-estimation/decoding method and describe how it can be mated with a soft decoder.
We also describe our EM-based method to learn the GMM channel parameters.
In \secref{bench}, we detail several benchmarks that will be used in our numerical comparisons, including Bussgang-linearized PBiGAMP and linear-MMSE symbol decoding with pilot-aided channel estimation.
In \secref{results}, we report numerical results, and in \secref{conc} we conclude. 

\emph{Notation}---%
We use 
boldface uppercase letters like $\boldsymbol{B}$ to denote matrices and 
boldface lowercase letters like $\boldsymbol{b}$ to denote vectors, 
where $b_i$ represents the $i$th element of $\boldsymbol{b}$, and 
$[\boldsymbol{B}]_{i,j}$ represents the $i$th row and $j$th column of $\boldsymbol{B}$.
Also, $\I_M$ is the $M\times M$ identity matrix, 
$\1_M$ is the $M$-length vector of ones, 
$\0_M$ is the $M$-length vector of zeros, 
$\Diag(\boldsymbol{b})$ is the diagonal matrix formed from the vector $\boldsymbol{b}$, 
$\diag(\boldsymbol{B})$ is the vector formed from the diagonal of matrix $\boldsymbol{B}$, 
$\F_N$ is the $N\times N$ unitary discrete Fourier transform (DFT) matrix,
$\F_N^{1:L}$ is the matrix formed by the first $L$ columns of $\F_N$, 
$\f_N^i$ is the $i$th column of $\F_N$, and 
$f_N^{ij}$ is the $(i\!+\!1,j\!+\!1)$th element of $\F_N$. 
For matrices and vectors, 
$(\cdot)\tran$ denotes transpose,
$(\cdot)\herm$ denotes conjugate transpose, 
$(\cdot)^*$ denotes conjugate, and
$\otimes$ denotes the Kronecker product.
Likewise, $\odot$, $\oslash$, and $|\cdot|^{\odot 2}$ denote element-wise multiplication, division, and absolute-value squared, respectively. 
Finally, the probability density function (pdf) of a multivariate complex Gaussian random vector $\x$ with mean $\hat{\x}$ and covariance $\boldsymbol{\Sigma}$ will be denoted by $\CN(\x;\hat{\x},\boldsymbol{\Sigma})$.

\section{System model} \label{sec:sysmodel}

\subsection{Single-Carrier Block Transmission Model} \label{sec:SC}

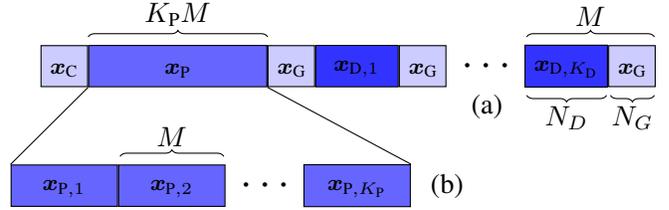
\begin{figure}
	\begin{tikzpicture}[scale=1.06, transform shape]
		\node (N_G1)[rectangle,draw, color= black, fill = blue!20!white, text width=0.35cm, text height=0.23cm]	at (-0.5,0) 
				{$\begin{footnotesize}\xCP \end{footnotesize}$};
  	        \node (N_P)[rectangle,draw, right=0cm of N_G1, color= black, fill = blue!60!white, text width=2.0cm, text height=0.24cm,align = center] 	
  	    		{$\begin{footnotesize} \xP \end{footnotesize}$};
		\node (N_G2)[rectangle,draw, right=0cm of N_P, color= black, fill = blue!20!white, text width=0.35cm, text height=0.23cm,align = center] 
				{$\begin{footnotesize} \x_{\textrm{G}} \end{footnotesize}$};
		\node (N_D1)[rectangle,draw, right=0cm of N_G2, color= black, fill = blue!80!white, text width=0.8cm, text height=0.20cm,align = center]	
				{$\begin{footnotesize} \x_{\textrm{D},1} \end{footnotesize}$};
		\node (N_G3)[rectangle,draw, right=0cm of N_D1, color= black, fill = blue!20!white, text width=0.35cm, text height=0.23cm,align = center]	
				 {$\begin{footnotesize} \x_{\textrm{G}} \end{footnotesize}$}; 
		\node (N_cdot)  [right=0cm of N_G3, scale=1.5, align=center]  {$\cdots$}; 	    
  	        \node (N_D2)[rectangle,draw, right=-0.1cm of N_cdot, color= black, fill = blue!80!white, text width=0.8cm, text height=0.20cm,align = center]	
				{$\begin{footnotesize} \x_{\textrm{D},\Kd} \end{footnotesize}$};
		\node (N_G4)[rectangle,draw, right=0cm of N_D2, color= black, fill = blue!20!white, text width=0.35cm, text height=0.23cm,align = center]	
				 {$\begin{footnotesize} \x_{\textrm{G}} \end{footnotesize}$};

  		\draw[decoration={brace,mirror,raise=5pt},decorate]
  					(5.3,-0.2) -- node[below=5pt] {$N_D$} (6.3,-0.2);
  		\draw[decoration={brace,mirror,raise=5pt},decorate]
  					(6.35,-0.2) -- node[below=5pt] {$N_G$} (6.9,-0.2);
  	        \draw[decoration={brace,raise=5pt},decorate]
  					(5.3,0.2) -- node[above=5pt] {$M$} (6.9,0.2);
  		\draw[decoration={brace,raise=5pt},decorate]
  					(-0.2,0.2) -- node[above=5pt] {$\Kp M$} (2.05,0.2);
		\draw (4.8,-0.5) node {(a)};

  		\node (N_PP1)[rectangle,draw, color= black, fill = blue!60!white, text width=1.1cm, text height=0.2cm,align = center] 	at (-0.5,-1.5) 
  	    		{$\begin{footnotesize} \x_{\textrm{P},1} \end{footnotesize} $};
		\node (N_PP2)[rectangle,draw, right=0cm of N_PP1, color= black, fill = blue!60!white, text width=1.1cm, text height=0.20cm,align = center] 	
 	    		{$\begin{footnotesize} \x_{\textrm{P},2} \end{footnotesize} $};	
		\node (N_cdot1)  [right=0cm of N_PP2, scale=1.5, align=center]  {$\cdots$};
		\node (N_PP3)[rectangle,draw, right=-0.1cm of N_cdot1, color= black, fill = blue!60!white, text width=1.1cm, text height=0.2cm,align = center] 	
  	    		{$\begin{footnotesize} \x_{\textrm{P},\Kp} \end{footnotesize} $};
		\draw[decoration={brace,raise=5pt},decorate]
  					(0.20,-1.3) -- node[above=5pt] {$M$} (1.5,-1.3);  

  		\draw[-] ($(N_P.south west)  +(-0.0,0.0)$) -- ($(N_PP1.north west)  +(0.0,0.0)$);
		\draw[-] ($(N_P.south east)  +(-0.0,0.0)$) -- ($(N_PP3.north east)  +(0.0,0.0)$);
		\draw (4.3,-1.5) node {(b)};  	    
  	    		
	\end{tikzpicture}
    \caption{(a) The transmission structure, containing cyclic-prefixed pilots $[\xCP,\xP]$ and data blocks $\x_{\textrm{D},k}$ separated by guard blocks $\xG$.
             (b) The block structure of the pilot sequence $\xP$.
             }
    \label{fig:block_struc}
\end{figure}

We consider a single-carrier block transmission system where the transmitted frame takes the form
\begin{align}
\tilde{\x} &= [\xP\tran, \xD\tran]\tran,
\end{align}
with $\xP$ a pilot frame and $\xD$ a data frame.
For compatibility with the IEEE 802.11ad standard \cite{802.11ad}, we assume that
the data frame consists of $\Kd$ guard-separated data blocks with guard length $\Ng$, and
the pilot frame consists of $\Kp$ pilot blocks with a cyclic-prefix (CP) structure.
In particular,
$\xD=[\xG\tran,\x_{\textrm{D},1}\tran,\xG\tran,
      \dots,
      \xG\tran,\x_{\textrm{D},\Kd}\tran,\xG\tran]\tran$,
where $\xG\in\Complex^{\Ng}$, 
$\x_{\textrm{D},k} \in \mathcal{S}^{\Nd}$, and
$\mathcal{S}$ is a $2^A$-ary complex symbol alphabet.
Note the CP structure induced by the guards.
Furthermore, we assume that
$\xP=[\xCP,\x_{\textrm{P},1}\tran,\dots,\x_{\textrm{P},\Kp}\tran]\tran$, where
the last $\Ncp$ elements of each $\x_{\textrm{P},k}\in\Complex^{M}$ 
equal $\xCP\in\Complex^{\Ncp}$,
so that the tail of each pilot block acts as the CP for the next block.
Finally, we assume that $M=\Nd+\Ng$. 
The assumed frame structure is illustrated in \figref{block_struc}(a).

The data sequences $\x_{D,k}$ are constructed as follows.
First, $\Nb$ information bits $\bb\defn[b_1,\dots,b_{\Nb}]\tran$ are coded and then interleaved, yielding the coded bits $\cc\in\{0,1\}^{A \Kd\Nd}$ and a 
code rate of $R=\frac{\Nb}{A \Kd \Nd}$.
Next, the coded bits are partitioned into $\Kd\Nd$ groups of $A$ bits, $\cc \defn [\cc_0\tran, \dots, \cc_{\Kd\Nd-1}\tran]\tran$, 
where each group $\cc_n \defn [c_{n,1},\dots,c_{n,A}]\tran$ determines the value of one data symbol.
By partitioning the $\Kd \Nd$ data symbols into $\Kd$ blocks of $\Nd$ symbols, one obtains the data sequences $\x_{\textrm{D},k}$ for $k=1,\dots,\Kd$. 

\subsection{Propagation and Few-Bit ADC Model} \label{sec:Quant}

The frame $\tilde{\x}$ is modulated using a square-root raised-cosine pulse, 
upconverted,
propagated through a noisy and frequency-selective channel (using possibly many antennas with analog beamforming at the transmitter and/or receiver),
downconverted, 
filtered with a square-root raised cosine pulse,
and sampled at the baud rate.
We will assume that the beamformed baseband channel impulse response,
$\h\defn[h_0, \dots, h_{L-1}]\tran$,
has length $L\leq \min\{\Ncp,\Ng\}-1$ and 
is invariant during the transmission of $\tilde{\x}$.
In this case, after discarding the received samples corresponding to the first $\xCP$ and $\xG$ sequences,
the unquantized received samples can be collected into the matrix 
\begin{align}
\U = \Hmat\X + \W 
\label{eq:unquant} ,
\end{align}
where $K\defn\Kp+\Kd$.
In \eqref{eq:unquant}, 
$\Hmat\in\Complex^{M\times M}$ is the circulant matrix with first column $[\h\tran~\0_{M-L}\tran]\tran$, 
$\W\in \Complex^{M\times K}$ contains additive white Gaussian noise (AWGN) with variance $\sigma_w^2$, which is assumed to be known,\footnote{\label{foot:noise}The noise variance could be estimated using the EM-PBiGAMP procedure described in \cite{Parker:JSTSP:16}, but we leave the verification of this approach to future work.  See \cite{Ziniel:TSP:15} for AWGN-variance learning under 1-bit quantization, referred to as the ``probit link'' in the context of binary classification.}
and the $k$th column of $\X\in\Complex^{M\times K}$ equals 
$\x_{\textrm{P},k}$ when $k\in\{1,\dots,\Kp\}$
or $[\x_{\textrm{D},k-\Kp}\tran,\xG\tran]\tran$ when $k>\Kp$.
Likewise, we can write \eqref{eq:unquant} in vectorized form as
\begin{align}
\uu = (\I_K \otimes \Hmat)\x +\w
\label{eq:vec_unquant} ,
\end{align}
with 
$\uu \defn \vvec(\U)$, 
$\x \defn \vvec(\X)$,
$\w \defn \vvec(\W)$, and
$\otimes$ denoting the Kronecker product.
It can be shown that $\x$ equals $\tilde{\x}$ with the first $\xCP$ and $\xG$ sequences removed.

The output of the few-bit ADC is modeled as
\begin{align}\label{eq:vec_quant}
\y = \mathcal{Q}\big(\uu\big),
\end{align}
where the quantization $\mathcal{Q}(\cdot)$ applies component-wise.
Although not required by our methodology, we will assume in our numerical experiments that $b$-bit uniform mid-rise quantization \cite{Max:IRETIT:60} is separately applied to the real and imaginary parts, i.e., 
\begin{align}\label{eq:quant_scalar}
y_m &= \text{sign}(\RRe{(u_m)})\left( \text{min}\left\{\left\lceil\frac{|\RRe{(u_m)}|}{\bigtriangleup_{\RRe}}\right\rceil,2^{b-1} \right\}-\frac{1}{2} \right) \\
  &~~+\text{j}\ \text{sign}(\IIm{(u_m)})\left( \text{min}\left\{\left\lceil\frac{|\IIm{(u_m)}|}{\bigtriangleup_{\IIm}}\right\rceil,2^{b-1} \right\}-\frac{1}{2} \right) 
\nonumber ,
\end{align}
where 
$\bigtriangleup_{\RRe} \triangleq \sqrt{\Ex \big[\RRe(u_m)^2 \big]}\bigtriangleup_b$,
$\bigtriangleup_{\IIm} \triangleq \sqrt{\Ex \big[\IIm(u_m)^2 \big]}\bigtriangleup_b$, 
and $\bigtriangleup_{\textrm{b}}$ is chosen to minimize the mean-squared error (MSE) $\Ex\big[|y_m-u_m|^2\big]$ under Gaussian $u_m$.
The average powers $\Ex \big[\RRe(u_m)^2 \big]$ and $\Ex \big[\IIm(u_m)^2 \big]$ can be measured by analog circuits before the ADC.
When $b\!>\!1$, such measurements are typically performed as part of automatic gain control.

\subsection{Channel Model for Propagation}\label{sec:60Gchannel}

For signal propagation, we used the 60~GHz wireless local area network (WLAN) channel model adopted by the IEEE~802.11ad task group \cite{Maltsev:Tech:10}, which was a result of extensive channel measurement studies in \cite{Maltsev:JSAC:09}.
It specifies that the continuous-space/time channel impulse response $h(t;\phi_{\text{tx}},\theta_{\text{tx}}, \phi_{\text{rx}},\theta_{\text{rx}})$, as a function of the lag $t$, the azimuth angles $(\phi_{\text{tx}},\phi_{\text{rx}})$, and the elevation angles $(\theta_{\text{tx}},\theta_{\text{rx}})$, takes the form%
\begin{subequations}
\label{eq:60Gchannel}
\begin{align}
\lefteqn{ h(t;\phi_{\text{tx}},\theta_{\text{tx}}, \phi_{\text{rx}},\theta_{\text{rx}}) }\nonumber\\
&= \sum_{i=1}^I \alpha^{(i)} C^{(i)}\Big( t-\tau^{(i)}; 
        \phi_{\text{tx}} - \Phi_{\text{tx}}^{(i)}, 
        \theta_{\text{tx}} - \Theta_{\text{tx}}^{(i)}, 
\nonumber\\ &\hspace{25mm} 
        \phi_{\text{rx}} - \Phi_{\text{rx}}^{(i)}, 
        \theta_{\text{rx}} - \Theta_{\text{rx}}^{(i)}\Big) \\
\lefteqn{ C^{(i)}(t;\phi_{\text{tx}},\theta_{\text{tx}}, \phi_{\text{rx}},\theta_{\text{rx}}) }\nonumber\\
&= \sum_{u=1}^{U^{(i)}} \alpha^{(i,u)} \delta(t-\tau^{(i,u)}) 
\delta( \phi_{\text{tx}} - \Phi_{\text{tx}}^{(i,u)})
\delta(\theta_{\text{tx}} - \Theta_{\text{tx}}^{(i,u)})
\nonumber\\&\hspace{24mm}\times 
\delta(\phi_{\text{rx}} - \Phi_{\text{rx}}^{(i,u)})
\delta(\theta_{\text{rx}} - \Theta_{\text{rx}}^{(i,u)}),
\end{align}
\end{subequations}
where 
\begin{itemize}
\item $\alpha^{(i)}$ and $C^{(i)}(t;\phi_{\text{tx}},\theta_{\text{tx}}, \phi_{\text{rx}},\theta_{\text{rx}})$ are the gain and channel impulse response of the $i$th cluster, respectively,
\item $\tau^{(i)}$, $\Phi_{\text{tx}}^{(i)}$, $\Theta_{\text{tx}}^{(i)}$, $\Phi_{\text{rx}}^{(i)}$, $\Theta_{\text{rx}}^{(i)}$ are the delay-angle coordinates of the $i$th cluster,
\item $\alpha^{(i,u)}$ is the gain of the $u$th ray of the $i$th cluster,
\item $\tau^{(i,u)}$, $\Phi_{\text{tx}}^{(i,u)}$, $\Theta_{\text{tx}}^{(i,u)}$, $\Phi_{\text{rx}}^{(i,u)}$, $\Theta_{\text{rx}}^{(i,u)}$ are the relative delay-angle coordinates of the $u$th ray of the $i$th cluster,
\item $I$ is the number of clusters and $U^{(i)}$ is the number of rays in the $i$th cluster, and
\item $\delta(\cdot)$ is the Dirac delta.
\end{itemize}
The discrete-time impulse response coefficients $\{h_l\}$ are constructed from $h(t;\phi_{\text{tx}},\theta_{\text{tx}}, \phi_{\text{rx}},\theta_{\text{rx}})$ via pulse-shaping and beamforming, i.e.,
\begin{align}
h_l
&= \int h(t;\phi_{\text{tx}},\theta_{\text{tx}}, \phi_{\text{rx}},\theta_{\text{rx}}) 
        g(lT-t)
\nonumber\\&\hspace{5mm}\times 
        b_{\text{tx}}(\phi_{\text{tx}},\theta_{\text{tx}})
        b_{\text{rx}}(\phi_{\text{rx}},\theta_{\text{rx}})
        \dif t
        \dif \phi_{\text{tx}}
        \dif \theta_{\text{tx}}
        \dif \phi_{\text{rx}}
        \dif \theta_{\text{rx}}
\label{eq:hl} ,
\end{align}
where 
$g(\cdot)$ is the pulse shape specified in the 802.11ad standard (i.e., raised-cosine with rolloff 0.25), 
$T$ is the baud interval,
and 
$b_{\text{tx}}(\phi_{\text{tx}},\theta_{\text{tx}})$ and 
$b_{\text{rx}}(\phi_{\text{rx}},\theta_{\text{rx}})$ are beam responses.

\begin{figure}
    \newcommand{\sz}{0.7}
    \psfrag{(a)}[b][b][\sz]{(a)}
    \psfrag{(b)}[b][b][\sz]{(b)}
    \psfrag{(c)}[b][b][\sz]{(c)}
	\psfrag{lag j}[][t][\sz]{\sf lag $l$}
	\psfrag{PDP}[b][b][\sz]{\sf PDP [dB]}
	\psfrag{amplitude}[b][b][\sz]{\sf amplitude}
    \centering
    \includegraphics[width=\figwid,clip]{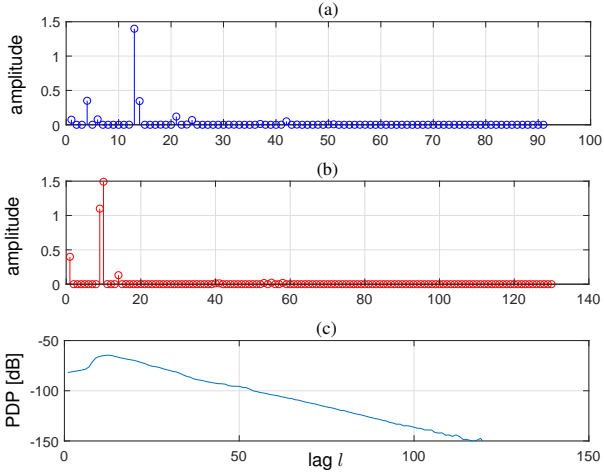}
    \caption{For the 802.11ad 60~GHz ``conference room'' channel, typical realizations of $|h_l|$ versus $l$ are shown in (a) and (b), and the power-delay profile is shown in (c).}
    \label{fig:chan_realization}
\end{figure}

Based on extensive physical channel measurements, 
statistical models for the 60GHz WLAN channel parameters were proposed in \cite{Maltsev:Tech:10}, and 
Matlab code to generate realizations from this model (including optimized analog beamforming) was provided in \cite{Maslennikov:Tech:10}.
Typical realizations of the resulting $\{|h_l|\}_{l=0}^{L-1}$ from the ``conference room'' environment are shown in Figs.~\ref{fig:chan_realization}(a)-(b), which show that the channel taps are approximately sparse.
The channel power-delay profile (PDP), $\Ex\{|h_l|^2\}$ versus $l$, is plotted in \figref{chan_realization}(c), with the expectation approximated by an average of 50\,000 realizations. 
There it can be seen that the PDP decays exponentially with lag $l$, i.e., the index into $\h$.

\subsection{Channel Model for Estimation} \label{sec:GMM}

The channel model as given in \eqref{eq:hl} is difficult to directly exploit for channel estimation. 
Therefore, for channel estimation, we propose to use a $D$-state Gaussian-mixture model (GMM) for the channel vector $\h$, as suggested in \cite{Schniter:JSTSP:11} for $D=2$. 
For general $D\geq 1$, the GMM specifies a pdf of the form 
\begin{subequations}
\label{eq:GMM} 
\begin{align}
p(\h;\blam,\bnu) 
&= \prod_{l=0}^{L-1} p(h_l;\blam_l,\bnu_l) \\
p(h_l;\blam_l,\bnu_l) 
&= \sum_{d=1}^D \lambda_{l,d} \,\CN(h_l;0,\nu_{l,d}) ,
\end{align}
\end{subequations}
where $\lambda_{l,d}\geq 0$ and $\nu_{l,d}>0$ are the weight and variance of the $d$th mixture component of the $l$ tap, and $\sum_{d=1}^D \lambda_{l,d}=1~\forall l$.
Also, $\blam_l\defn[\lambda_{l,1},\dots,\lambda_{l,D}]\tran$ and $\blam\defn[\blam_0\tran,\dots,\blam_{L-1}\tran]\tran$, with similar definitions for $\bnu_l$ and $\bnu$.
In principle, the GMM parameters, $\blam$ and $\bnu$, could be empirically estimated from a corpus of training data using the standard EM-based approach to fitting a GMM \cite[p. 435]{Bishop:Book:07}. 
As an alternative, these parameters can be estimated online from the quantized measurements $\y$ using the EM-AMP-based method described in \secref{EM}.


\section{Turbo Equalization with PBiGAMP} \label{sec:JCEDpbigamp}

Our principle goal is to infer the information bits $\bb$ from the few-bit measurements $\y$
under the block-transmission model from \secref{SC},
the few-bit ADC model from \secref{Quant},
and the GMM channel model from \secref{GMM}. 
In particular, we aim to compute the marginal posterior probabilities $\{p(b_i|\y)\}_{i=1}^{\Nb}$, which can be decomposed as
\begin{align}
\lefteqn{ p(b_i|\y) 
= \sum_{\bb_{-i}} p(\bb|\y) 
= \sum_{\bb_{-i}} \frac{p(\y|\bb) p(\bb)}{p(\y)}
\propto \sum_{\bb_{-i}} p(\y|\bb) }
\label{eq:fg1}\\
&=\sum_{\bb_{-i},\x,\cc} \int_{\Complex^L} 
p(\y|\h,\x) p(\h) p(\x|\cc) p(\cc|\bb) \dif\h
\label{eq:fg2}\\
&=\sum_{\bb_{-i},\cc} p(\cc|\bb) 
\sum_{\x} \int_{\Complex^L} \Bigg[\prod_{m=1}^{MK}p(y_m|\h,\x) \Bigg]\Bigg[ \prod_{l=0}^{L-1} p(h_l) \Bigg] \dif\h
\nonumber\\&\quad\times
\Bigg[\prod_{k=1}^{\Kd} \prod_{n=0}^{\Nd-1} p(x_{(\Kp+k-1)M+n}|\cc_{(k-1)\Nd+n}) \Bigg]
\label{eq:fg} ,
\end{align}
for $\bb_{-i}\triangleq[b_1,\dots,b_{i-1},b_{i+1},\dots,b_{\Nb}]\tran$.
Above, \eqref{eq:fg1} is due to Bayes rule and the assumption that the information bits $\bb$ are uniformly distributed;
\eqref{eq:fg2} is due to the dependency relationships among the random vectors $\y$, $\h$, $\x$, $\cc$, and $\bb$;
and \eqref{eq:fg} is due to the separable nature of $ p(\y|\h,\x)$, $p(\h)$, and $p(\x|\cc)$. 
In particular, the pmfs $p(x_{(\Kp+k-1)M+n}|\cc_{(k-1)\Nd+n})$ for $k=1,\dots,\Kd$ and $n=0,\dots,\Nd-1$ are determined by the bit-to-symbol mapping,
and the likelihood function $p(y_m|\h,\x)$ can be obtained from \eqref{eq:vec_unquant}-\eqref{eq:vec_quant}. 
Details are provided in the sequel.

\begin{figure}
	\begin{tikzpicture}[scale=1.28, transform shape]
  	   \node (fb1)[rectangle,draw,fill=black!100] at (0,0) {};
  	   \node (fb2)[below =  0.8cm of fb1, rectangle,draw,fill=black!100]  {};
  	   \node (fb3)[below =  0.8cm of fb2, rectangle,draw,fill=black!100]  {};
  	   \node (b1) [right =  0.4cm of fb1, circle,draw, minimum size = 0.01cm,label=above:{$b_1$}]  {};
  	   \node (b2) [right =  0.4cm of fb2, circle,draw, minimum size = 0.01cm,label=above:{$b_2$}]  {};
	   \node (b3) [right =  0.4cm of fb3, circle,draw, minimum size = 0.01cm,label=above:{$b_3$}]  {};  	   
       \draw [fill=black, label=above:{$p(\cc|\bb)$}]     ($(b1.north east)+(+0.4,0.35)$) rectangle ($(b3.south east)+(+0.65,-0.35)$);   	
       \node (c1) [ circle,draw, minimum size = 0.01cm,label={[xshift =-0.6cm, yshift=0.3cm]right:$c_{1,1}$}] at (2.05,0.2) {};   
       \node (c2) [below =  0.5cm of c1, circle,draw, minimum size = 0.01cm,label={[xshift =-0.6cm, yshift=0.3cm]right:$c_{1,2}$}]  {}; 
       \node (c3) [below =  0.5cm of c2, circle,draw, minimum size = 0.01cm,label={[xshift =-0.6cm, yshift=0.3cm]right:$c_{2,1}$}]  {};   
       \node (c4) [below =  0.5cm of c3, circle,draw, minimum size = 0.01cm,label={[xshift =-0.6cm, yshift=0.3cm]right:$c_{2,2}$}]  {};   
  	   
  	   \node (M1) [right =  0.4cm of c1,rectangle,draw,label=above:{},pattern=north west lines,label={[xshift =-0.4cm,yshift=0.3cm]right:$\mathsmaller{\mathcal{M}_0}$}]{}; 
  	   \node (M2) [right =  0.4cm of c2,rectangle,draw,label={[xshift =-0.4cm,yshift=0.3cm]right:$\mathsmaller{\mathcal{M}_1}$},pattern=north west lines,]  {};
  	   \node (M3) [right =  0.4cm of c3,  rectangle,draw,label={[xshift =-0.4cm,yshift=0.3cm]right:$\mathsmaller{\mathcal{M}_2}$},fill=black!100]  {};
  	   \node (M4) [right =  0.4cm of c4,  rectangle,draw,label={[xshift =-0.4cm,yshift=0.3cm]right:$\mathsmaller{\mathcal{M}_3}$},fill=black!100]  {};

  	   \node (x1) [right =  0.4cm of M1, circle,draw, minimum size = 0.01cm,label={[xshift =-0.4cm, yshift=0.3cm]right:$x_0$}, dashed]  {};
  	   \node (x2) [right =  0.4cm of M2, circle,draw, minimum size = 0.01cm,label={[xshift =-0.4cm, yshift=0.3cm]right:$x_1$}, dashed]  {};
  	   \node (x3) [right =  0.4cm of M3, circle,draw, minimum size = 0.01cm,label={[xshift =-0.4cm, yshift=0.3cm]right:$x_2$}]  {};
  	   \node (x4) [right =  0.4cm of M4, circle,draw, minimum size = 0.01cm,label={[xshift =-0.4cm, yshift=0.3cm]right:$x_3$}]  {};

  	   \node (py1) [right =  0.8cm of x1, rectangle,draw,label=above:{$y_0$},fill=black!100]    {};
  	   \node (py2) [right =  0.8cm of x2, rectangle,draw,label=above:{$y_1$},fill=black!100]   {};
  	   \node (py3) [right =  0.8cm of x3, rectangle,draw,label=above:{$y_2$},fill=black!100]   {};
  	   \node (py4) [right =  0.8cm of x4, rectangle,draw,label=above:{$y_3$},fill=black!100]   {};
  	   
  	   \node (h1) [ circle,draw, minimum size = 0.01cm,label={[xshift =-0.4cm, yshift=0.3cm]right:$h_0$}] at (5.5,-0.2) {};
 	   \node (h2) [below =  0.48cm of h1, circle,draw, minimum size = 0.01cm,label={[xshift =-0.4cm, yshift=0.3cm]right:$h_1$}]  {};
 	   \node (h3) [below =  0.48cm of h2, circle,draw, minimum size = 0.01cm,label={[xshift =-0.4cm, yshift=0.3cm]right:$h_2$}]  {};
 	   
       \node (pd1) [right =  0.4cm of h1, rectangle,draw,label=above:{},fill=black!100]   {};   
       \node (pd2) [right =  0.4cm of h2, rectangle,draw,label=above:{},fill=black!100]   {};   
       \node (pd3) [right =  0.4cm of h3, rectangle,draw,label=above:{},fill=black!100]   {};  
       
	  \path[-] (fb1) edge node {} (b1);
	  \draw[-] ($(b1.east)  +(0.0,0.0)$) -- ($(b1.east)  +(0.5,0.0)$);
	  \path[-] (fb2) edge node {} (b2);
	  \draw[-] ($(b2.east)  +(0.0,0.0)$) -- ($(b2.east)  +(0.5,0.0)$);
	  \path[-] (fb3) edge node {} (b3);
	  \draw[-] ($(b3.east)  +(0.0,0.0)$) -- ($(b3.east)  +(0.5,0.0)$);
	   
	   \draw[-] ($(c1.west)  +(-0.5,0.0)$) -- ($(c1.west)  +(0.0,0.0)$);	   
	   \draw[-] ($(c2.west)  +(-0.5,0.0)$) -- ($(c2.west)  +(0.0,0.0)$);
	   \draw[-] ($(c3.west)  +(-0.5,0.0)$) -- ($(c3.west)  +(0.0,0.0)$);
	   \draw[-] ($(c4.west)  +(-0.5,0.0)$) -- ($(c4.west)  +(0.0,0.0)$);
	   
	   \draw[-] ($(c1.east)  +(-0.0,0.0)$) -- ($(M3.west)  +(0.0,0.0)$);
	   \draw[-] ($(c2.east)  +(-0.0,0.0)$) -- ($(M3.west)  +(0.0,0.0)$);
	   \draw[-] ($(c3.east)  +(-0.0,0.0)$) -- ($(M4.west)  +(0.0,0.0)$);
	   \draw[-] ($(c4.east)  +(-0.0,0.0)$) -- ($(M4.west)  +(0.0,0.0)$);
	   
	   \draw[-] ($(M1.east)  +(-0.0,0.0)$) -- ($(x1.west)  +(0.0,0.0)$);
	   \draw[-] ($(M2.east)  +(-0.0,0.0)$) -- ($(x2.west)  +(0.0,0.0)$);
	   \draw[-] ($(M3.east)  +(-0.0,0.0)$) -- ($(x3.west)  +(0.0,0.0)$);
	   \draw[-] ($(M4.east)  +(-0.0,0.0)$) -- ($(x4.west)  +(0.0,0.0)$);
	   
	   \draw[-] ($(x1.east)  +(-0.0,0.0)$) -- ($(py1.west)  +(0.0,0.0)$);
	   \draw[-] ($(x1.east)  +(-0.0,0.0)$) -- ($(py2.west)  +(0.0,0.0)$);
	   \draw[-] ($(x2.east)  +(-0.0,0.0)$) -- ($(py1.west)  +(0.0,0.0)$);
	   \draw[-] ($(x2.east)  +(-0.0,0.0)$) -- ($(py2.west)  +(0.0,0.0)$);
	   \draw[-] ($(x3.east)  +(-0.0,0.0)$) -- ($(py3.west)  +(0.0,0.0)$);
	   \draw[-] ($(x3.east)  +(-0.0,0.0)$) -- ($(py4.west)  +(0.0,0.0)$);
	   \draw[-] ($(x4.east)  +(-0.0,0.0)$) -- ($(py3.west)  +(0.0,0.0)$);
	   \draw[-] ($(x4.east)  +(-0.0,0.0)$) -- ($(py4.west)  +(0.0,0.0)$);
	   
	   \draw[-] ($(py1.east)  +(-0.0,0.0)$) -- ($(h1.west)  +(0.0,0.0)$);
	   \draw[-] ($(py1.east)  +(-0.0,0.0)$) -- ($(h2.west)  +(0.0,0.0)$);
	   \draw[-] ($(py1.east)  +(-0.0,0.0)$) -- ($(h3.west)  +(0.0,0.0)$);
	   \draw[-] ($(py2.east)  +(-0.0,0.0)$) -- ($(h1.west)  +(0.0,0.0)$);
	   \draw[-] ($(py2.east)  +(-0.0,0.0)$) -- ($(h2.west)  +(0.0,0.0)$);
	   \draw[-] ($(py2.east)  +(-0.0,0.0)$) -- ($(h3.west)  +(0.0,0.0)$);
	   \draw[-] ($(py3.east)  +(-0.0,0.0)$) -- ($(h1.west)  +(0.0,0.0)$);
	   \draw[-] ($(py3.east)  +(-0.0,0.0)$) -- ($(h2.west)  +(0.0,0.0)$);
	   \draw[-] ($(py3.east)  +(-0.0,0.0)$) -- ($(h3.west)  +(0.0,0.0)$);
	   \draw[-] ($(py4.east)  +(-0.0,0.0)$) -- ($(h1.west)  +(0.0,0.0)$);
	   \draw[-] ($(py4.east)  +(-0.0,0.0)$) -- ($(h2.west)  +(0.0,0.0)$);
	   \draw[-] ($(py4.east)  +(-0.0,0.0)$) -- ($(h3.west)  +(0.0,0.0)$);
	   
	   \draw[-] ($(h1.east)  +(-0.0,0.0)$) -- ($(pd1.west)  +(0.0,0.0)$);
	   \draw[-] ($(h2.east)  +(-0.0,0.0)$) -- ($(pd2.west)  +(0.0,0.0)$);
	   \draw[-] ($(h3.east)  +(-0.0,0.0)$) -- ($(pd3.west)  +(0.0,0.0)$);
	   
	   \draw[line width=0.07mm, dashed]     ($(c1.north west)+(-2.3,0.35)$) rectangle ($(c4.south west)+(0.43,-0.25)$); 
	   \draw[line width=0.07mm, dashed]     ($(x1.north west)+(-0.83,0.35)$) rectangle ($(py4.south east)+(1.75,-0.25)$);   	   
	   
	   \node[below left = 0.02cm and 0.7cm of c4, scale=0.5]  {soft decoding};
	   \node[below left = 0.02cm and -1.7cm of py4, scale=0.5]  {soft equalization with an unknown channel};
	   \node[above left = 0.4cm and 2.3cm of M1, scale=0.5, align=center]  {uniform \\[-0.5em] prior};
	   \node[above left = 0.4cm and 0.85cm of M1, scale=0.5, align=center]  {coding \&\\[-0.5em] interleaving};
	   \node[above left = 0.4cm and -0.5cm of M1, scale=0.5, align=center]  {symbol\\[-0.5em] mapping};	  
       \node[above right = 0.4cm and 1.2cm of M1, scale=0.5, align=center]  {observation\\[-0.5em] likelihood};		  
	   \node[above right = 0.38cm and 3cm of M1, scale=0.5, align=center]  {GMM\\[-0.5em] prior};	
	\end{tikzpicture}
    \caption{The factor graph corresponding to a toy example with 
$\Nb=3$ information bits $\{b_i\}$, 
$4$ interleaved/coded bits $\{c_{n,a}\}$, 
$A=2$ bits/symbol, 
$\Nd=2$ data symbols per block, 
$\Ng=0$ guard symbols per block, 
$\Kp=1$ pilot blocks, 
$\Kd=1$ data blocks, 
block length $M=\Nd+\Ng=2$, 
pilot symbols $x_0$ and $x_1$, 
data symbols $x_2$ and $x_3$, 
and $L=3$ channel taps. 
The node $y_m$ represents $p(y_m|z_m)$ and 
the node $\mathcal{M}_n$ represents the bit-to-symbol mapping for data symbols or the indicator pmf for pilot symbols.}
	\label{fig:fg}
\end{figure}
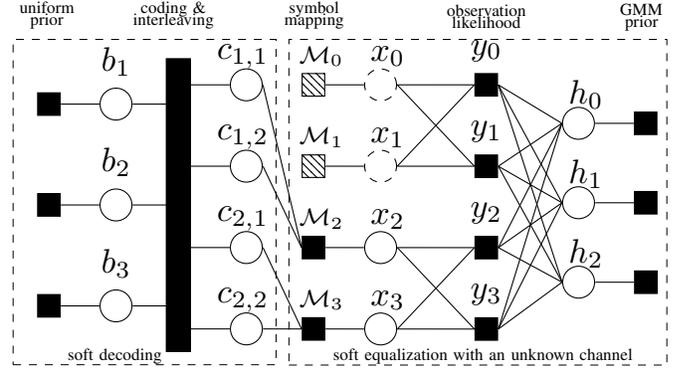

The structure in \eqref{eq:fg} can be visualized using the bipartite factor graph shown in \figref{fg}, where 
the solid rectangles represent the pdf factors and the open circles represent the variable nodes.
We find it convenient to partition the factor graph into two subgraphs:
the left subgraph corresponds to soft decoding and
the right subgraph corresponds to soft equalization with an unknown channel. 

\subsection{Belief Propagation} \label{sec:BP}

The posterior bit marginals $\{p(b_i|\y)\}_{i=1}^{\Nb}$ can in principle be computed from \eqref{eq:fg}, but doing so is impractical from the standpoint of complexity.
A practical alternative is to perform belief-propagation (BP) using the sum-product algorithm (SPA) \cite{Kschischang:TIT:01}, which passes messages along the edges of the factor graph in \figref{fg}.
For discrete-valued variables like $b_i, c_{n,a}, x_n$, these messages come in the form of pmfs,
while for continuous variables like $h_l$, these messages come in the form of pdfs.
When there are no loops (i.e., cycles) in the factor graph, BP computes exact marginals.
But \figref{fg} has loops, and so BP computes only approximate marginals. 
This is to be expected, given that exact inference in loopy graphs is NP hard \cite{Cooper:AI:90}.
Still, loopy BP often gives very good results, and so it has become popular for, e.g.,
turbo decoding, 
LDPC decoding, 
turbo equalization, 
inference of Markov random fields, 
multiuser detection, 
and compressive sensing. 

Exact implementation of the SPA is intractable for the soft-equalization subgraph in \figref{fg}. 
For exact SPA, the messages in and out of the $h_l$ nodes would take the form of Gaussian mixtures, with a mixture order that grows exponentially in the iterations.
As an alternative, one might consider passing only Gaussian approximations of these problematic SPA messages, an approach known as expectation propagation (EP) \cite{Minka:Diss:01}.
But since there are $MKL$ edges between the $\{h_l\}$ and $\{y_m\}$ nodes in \figref{fg}, the per-symbol complexity of EP would be $O(L)$, which contrasts with the $O(\log L)$ complexity of FFT processing.
Also, the fixed-points of EP are generally not well understood. 

\subsection{Background on PBiGAMP}\label{sec:pbigamp}

We now briefly provide some background on PBiGAMP, since many readers may not be familiar with the algorithm.
PBiGAMP \cite{Parker:JSTSP:16} is a computationally efficient approach to approximating the marginal posteriors of independent random variables $\{\rvx_n\}_{n=0}^{N-1}$ and $\{\rvh_l\}_{l=0}^{L-1}$ from measurements $\y=[y_0,\dots,y_{P-1}]\tran$ generated under a likelihood of the form
\begin{subequations}\label{eq:pbigamp}
\begin{align} 
p_{\rvyy|\rvzz}(\y|\z) &= \prod_{m=0}^{P-1} p_{\rvy_m|\rvz_m}(y_m|z_m) \\
\rvz_m &= \sum_{n=0}^{N-1} \sum_{l=0}^{L-1} \rvx_n z_m^{(n,l)} \rvh_l ,
\end{align}
\end{subequations}
where $z_m^{(n,l)}$ are known parameters. 
Throughout this subsection, we typeset random variables in san-serif font (e.g., $\rvy_m$) and non-random variables in serif font (e.g., $y_m$) for clarity.
Note that, in \eqref{eq:pbigamp}, $\rvz_m$ can be interpreted as noiseless bilinear measurements of the random vectors $\rvxx\triangleq[\rvx_0,\dots,\rvx_{N-1}]\tran$ and $\rvhh\triangleq[\rvh_0,\dots,\rvh_{L-1}]\tran$, and $p_{\rvy_m|\rvz_m}(y_m|z_m)$ can be interpreted as a noisy measurement channel.
Applications of \eqref{eq:pbigamp} include matrix compressive sensing, self-calibration, blind deconvolution, and joint channel/symbol estimation.

The PBiGAMP algorithm from \cite{Parker:JSTSP:16} is summarized in \tabref{pbigamp}.
There, the priors on $\rvx_n$ and $\rvh_l$ are denoted by $p_{\rvx_n}(x_n)$ and $p_{\rvh_l}(h_l)$, respectively. 
The approximate marginal posteriors, denoted by $p_{\rvx_n|\rvq_n}(x_n|\hat{q}_n;\nuq_n)$ and $p_{\rvh_l|\rvr_l}(h_l|\hat{r}_l;\nur_l)$, are specified in lines (D2)-(D3). 
Here, $\hat{q}_n,\nuq_n,\hat{r}_l,\nur_l$ are quantities computed iteratively by PBiGAMP.

\putTable{pbigamp}{The Scalar-Variance PBiGAMP Algorithm from \cite{Parker:JSTSP:16}
}{\scriptsize 
\begin{equation*}
\begin{array}{| r@{\,}c@{\,}l@{\,}r |}\hline 

 \multicolumn{3}{|l}{\textrm{Definitions:}}&\\[-1mm]
 ~p_{\rvz_m|\rvp_{m}\!}\big(z\giv\hat{p};\nup\big) 
 &\triangleq& \frac{p_{\rvy_{m}|\rvz_{m}\!}(y_{m} \giv z) \, \CN(z;\hat{p},\nup)}{\int p_{\rvy_{m}|\rvz_{m}\!}(y_{m} \giv z') \, \CN(z';\hat{p},\nup) \dif z'}    &\text{(D1)}\\ 
 p_{\rvh_{l}|\rvr_{l}\!}(h\giv\hat{r};\nur) 
	&\triangleq& \frac{p_{\rvh_{l}\!}(h) \, \CN(\hat{r};h,\nur)}{\int p_{\rvh_{l}\!}(h') \, \CN(\hat{r};h',\nur) \dif h'}&\text{(D2)}\\
 p_{\rvx_{n}|\rvq_{n}\!}(x\giv\hat{q};\nuq) 
	&\triangleq& \frac{p_{\rvx_{n}\!}(x) \, \CN(\hat{q};x,\nuq)}{\int p_{\rvx_n\!}(x') \, \CN(\hat{q};x',\nuq) \dif x'}&\text{(D3)}\\

 \multicolumn{3}{|l}{\textrm{Initialization:}}&\\
    \forall m:
      \hat{s}_{m}[0] &=& 0  & \text{(I1)}\\
    \forall n,l: \textrm{choose~} &
      \multicolumn{2}{l}{\hat{x}_{n}[1], \nux[1], \hat{h}_{l}[1], \nuh[1]} &\text{(I2)}\\

 \multicolumn{3}{|l}{\textrm{For $t=1,\dots T_{\max}$}}&\\
    \forall n: \hat{\z}^{(n,*)}{[t]}
   		&=& \sum_{l=0}^{L-1} \z^{(n,l)} \hat{h}_{l}[t] & \text{(R1)}\\[1mm]
    
    \forall l: \hat{\z}^{(*,l)}{[t]}
   		&=& \sum_{n=0}^{N-1} \hat{x}_{n}[t] \z^{(n,l)} & \text{(R2)}\\[1mm]  
    
    \hat{\z}^{(*,*)}{[t]}
   		&=& \sum_{n=0}^{N-1} \hat{x}_{n}[t] \hat{\z}^{(n,*)}{[t]} 
        \text{~or~} \sum_{l=0}^{L-1} \hat{h}_{l}[t]\hat{\z}^{(*,l)}{[t]} & 
        	\text{(R3)}\\[1mm]  
    
    \barnup[t]
  		&=& \frac{1}{P}\big( \nux[t] \sum_{n=0}^{N-1} \| \hat{\z}^{(n,*)}{[t]}\|^2 
                    &\\&& 
                    +\nuh[t] \sum_{l=0}^{L-1} \|\hat{\z}^{(*,l)}{[t]}\|^2 \big) &
  			\text{(R4)}\\[1mm]
    
    \nup[t]
   		&=& \barnup[t]+  \nux[t]\nuh[t]  
                    \frac{1}{P}\sum_{n=0}^{N-1} \sum_{l=0}^{L-1} 
   		    \|\z^{(n,l)}{[t]}\|^2 & \text{(R5)}\\[1mm]   		 
    
    \hat{\p}[t] 
   		&=& \hat{\z}^{(*,*)}{[t]} - \hat{\s}[t\!-\!1]\barnup[t]& 
   		    \text{(R6)}\\[1mm] 
    
    \nuz[t] 
   		&=& \frac{1}{P}\sum_{m=0}^{P-1}\var\{\rvz_{m}\giv\rvp_{m}\!=\!\hat{p}_{m}[t];\nup[t]\} & 
   			\text{(R7)}\\[1mm] 
    
    \forall m: \hat{z}_{m}[t] 
   		&=& \Ex[\rvz_{m}\giv\rvp_{m}\!=\!\hat{p}_{m}[t];\nup[t]] & 
   		 	\text{(R8)}\\[1mm] 
    
    \nus[t] 
   		&=& (1 -  \nuz[t]/\nup[t] )/\nup[t]  & \text{(R9)}\\[1mm] 		
    
    \hat{\s}[t] 
   		&=& ( \hat{\z}[t] - \hat{\p}[t])/\nup[t] & \text{(R10)}\\[1mm] 
    
    \nur[t]
   		&=&  \big( \nus[t]  \frac{1}{L}\sum_{l=0}^{L-1} \|\hat{\z}^{(*,l)}{[t]}\|^2 \big)^{-1} & 
   			\text{(R11)}\\[1mm]    	
    
    \forall l: \hat{r}_{l}[t] 
   		&=& \hat{h}_{l}[t] + \nur[t]\hat{\z}^{(*,l)\textsf{H}}[t]\hat{\s}[t] &\\
   		&&  -\nur[t]\nus[t]\nux[t]\hat{h}_{l}[t] \sum_{n=0}^{N-1}  
                \|\z^{(n,l)}\|^2 & \text{(R12)}\\[1mm]
    
    \nuq[t] 
   		&=& \big( \nus[t] \frac{1}{N}\sum_{n=0}^{N-1} \| \hat{\z}^{(n,*)}{[t]}\|^2 \big)^{-1} & 
   		\text{(R13)}\\[1mm]   		
    
    \forall n: \hat{q}_{n}[t] 
   		&=& \hat{x}_{n}[t] + \nuq[t] \hat{\z}^{(n,*)\textsf{H}}[t]\hat{\s}[t] &\\
   		 	&& -\nuq[t]\nus[t]\nuh[t] \hat{x}_{n}[t] \sum_{l=0}^{L-1} 
   		 		\|\z^{(n,l)}{}\|^2  &\text{(R14)}\\[1mm]	
   		 		
    \nuh[t\!+\!1] 
   		&=& \frac{1}{L}\sum_{l=0}^{L-1} \var\{\rvh_{l}\giv \rvr_l\!=\!\hat{r}_{l}[t]; \nur[t]\} & 	
   			\text{(R15)}\\[1mm] 
    
    \forall l: \hat{h}_{l}[t\!+\!1]    
   		&=& \Ex[\rvh_l \giv \rvr_l \!=\!\hat{r}_{l}[t]; \nur[t]]& \text{(R16)}\\[1mm] 
    
    \nux[t\!+\!1] 
   		&=& \frac{1}{N}\sum_{n=0}^{N-1}\var\{\rvx_{n}\giv\rvq_{n}\!=\!\hat{q}_{n}[t]; \nuq[t]\} & 
   		 	\text{(R17)}\\[1mm] 
    
    \forall n: \hat{x}_{n}[t\!+\!1] 
   		&=& \Ex[\rvx_{n}\giv \rvq_{n}\!=\!\hat{q}_{n}[t]; \nuq[t]]& \text{(R18)}\\[1mm] 	
 \multicolumn{3}{|l}{\textrm{end}}&\\\hline
\end{array}
\end{equation*}
}

In \cite{Parker:JSTSP:16}, PBiGAMP was derived as a computationally efficient approximation of the SPA for the likelihood model \eqref{eq:pbigamp}, assuming that $z_m^{(n,l)}$ are independent realizations of a zero-mean Gaussian random variable. 
This approximation is, in fact, exact in the large-system limit (i.e., $P,N,L\rightarrow\infty$ with fixed $N/P$ and $L/P$).
In \cite{Schulke:PRE:16}, PBiGAMP was analyzed using the replica method from statistical physics.
There it was shown that the large-system-limit performance of PBiGAMP can be accurately predicted by a scalar state-evolution.
For the case of i.i.d.\ Bernoulli-Gaussian $x_n$ and $h_l$, this state evolution was studied in detail and found to exhibit a sharp ``phase-transition'' behavior. 
Moreover, for certain combinations of measurement rates (i.e., $N/P$ and $L/P$) and sparsity rates on $x_n$ and $h_l$, PBiGAMP was shown to converge to the MMSE estimates of $\x$ and $\h$.
For other, more difficult, combinations of measurement and sparsity rates, PBiGAMP may not yield accurate estimates.
However, it is conjectured that no other polynomial-time method will yield accurate estimates in that case \cite{Schulke:PRE:16}.

\subsection{Soft Equalization via PBiGAMP} \label{sec:joint_pbigamp}

In this section, we describe how PBiGAMP can be applied to soft equalization of SC block transmissions over unknown FS channels measured by few-bit ADCs.

We begin by adapting the PBiGAMP likelihood model \eqref{eq:pbigamp} to the few-bit SC block-transmission model \eqref{eq:vec_unquant}-\eqref{eq:vec_quant}. 
First, we write the circulant channel matrix as $\Hmat=\sum_{l=0}^{L-1}h_l \J_l$, where $\J_l\in\Real^{M\times M}$ is the $l$-circulant delay matrix.
Then \eqref{eq:vec_quant} becomes
\begin{align}\label{eq:zm_pib}
y_m = \mathcal{Q}\left( \sum_{l=0}^{L-1}\sum_{n=0}^{MK-1} h_l [\I_K\otimes\J_l]_{m,n}x_n + w_m \right),
\end{align}
where $[\cdot]_{m,n}$ extracts the $m$th row and $n$th column of its matrix argument.
From \eqref{eq:pbigamp} and \eqref{eq:zm_pib}, we can readily identify the PBiGAMP quantities 
\begin{align} 
z_m^{(n,l)} 
&= [\I_K\otimes\J_l]_{m,n}  
\label{eq:zmnl} \\
p_{\rvy_m|\rvz_m}(y_m|z_m) 
&\defn \Pr\{y_m=\mathcal{Q}(z_m+\rvw_m)\} \\
&= \int_{\mathcal{Q}^{-1}(y_m)}\mathcal{CN}\big(w;z_m,\sigma_w^2) \dif w
\label{eq:pyzm} ,
\end{align}
where $\mathcal{Q}^{-1}(y_m)\subset \Complex$ is the region quantized to $y_m$.
We also identify the PBiGAMP dimensions $P=N=MK$.

For PBiGAMP's prior on $\rvh_l$, we assign the GMM from \eqref{eq:GMM}. 
For PBiGAMP's prior on $\rvx_n$, we treat the indices $n$ of data symbols differently from those of pilot and guard symbols.  
For the data indices $n \in \{(\Kp+k-1)M,\dots,(\Kp+k-1)M+\Nd-1\}_{k=1}^{\Kd}$, we assign
\begin{align} \label{eq:xn}
p_{\rvx_n}(x_n)
&= \sum_{j=1}^{2^A} \gamma_{n,j} \delta(x_n-s^{(j)}) ,
\end{align}
where 
$\delta(\cdot)$ is the Dirac delta, 
$\{s^{(1)},\dots,s^{(2^A)}\}\defn\mathcal{S}$ is the data-symbol alphabet, and
$\gamma_{n,j}=\Pr\{\rvx_n\!=\!s^{(j)}\}$ is the prior data-symbol pmf, which depends on the decoder outputs as described below. 
For pilot indices $n=0,\dots,\Kp M-1$ and guard indices $n\in\{(\Kp+k-1)M+\Nd,\dots, (\Kp+k)M-1\}_{k=1}^{\Kd}$, we assign the trivial prior $p_{\rvx_n}(x)=\delta(x-x_n)$ because the pilots and guards take on known deterministic values. 
Note that, although the data symbols $\rvx_n$ are discrete, PBiGAMP treats them as continuous random variables in $\Complex$.

The data-symbol pmf $\{\gamma_{n,j}\}_{j=1}^{2^A}$ is determined by the coded-bit priors $\Pr\{\rvc_{n,a}=c_a^{(j)}\}$ coming from the soft decoder, i.e.,
\begin{align}
\gamma_{n,j} 
&\defn \Pr\{\rvx_n\!=\!s^{(j)}\}
= \sum_{j'=1}^{2^A} \Pr\{\rvx_n\!=\!s^{(j)},\rvcc_n=\cc^{(j')}\} 
        \label{eq:s_prior0}\\
&= \sum_{j'=1}^{2^A} \underbrace{\Pr\{\rvx_n\!=\!s^{(j)}|\rvcc_n=\cc^{(j')}\}}_{\displaystyle \delta_{j-j'}} \Pr\{\rvcc_n=\cc^{(j')}\}\\
&= \Pr\{\rvcc_n=\cc^{(j)}\}
= \prod_{a=1}^A \Pr\{\rvc_{n,a}=c_a^{(j)}\} 
        \label{eq:s_prior},
\end{align}
where $\cc^{(j)}=[c_1^{(j)},\dots,c_A^{(j)}]\tran\in\{0,1\}^A$ 
is the coded-bit sequence corresponding to the symbol value $s^{(j)}$,
and $\delta_j$ is the Kronecker delta sequence.

We are now ready to apply PBiGAMP from \tabref{pbigamp}.
In the sequel, we omit the iteration index ``$[t]$'' for brevity.
From \eqref{eq:zmnl} and $\z^{(n,l)}\defn[z_0^{(n,l)},\dots,z_{MK-1}^{(n,l)}]\tran$, lines (R1)-(R3) of \tabref{pbigamp} become 
\begin{align}
\hat{\z}^{(n,*)} &= \sum_{l=0}^{L-1}\hat{h}_l[\I_K\otimes\J_l]_{:,n} = [\I_K\otimes\hat{\Hmat}  ]_{:,n} \label{eq:zns}  \\
\hat{\z}^{(*,l)} &= \sum_{n=0}^{MK-1}\hat{x}_n[\I_K\otimes\J_l]_{:,n}= \vvec\big(\J_l\hat{\X} \big)\label{eq:zsl}\\
\hat{\z}^{(*,*)} &= \sum_{l=0}^{L-1}\hat{h}_l\ \vvec\big(\J_l\hat{\X} \big) = \vvec\big( \hat{\Hmat} \hat{\X}  \big),  \label{eq:zss} 
\end{align}
where $[\cdot]_{:,n}$ extracts the $n$th column of its matrix argument,
$\hat{\Hmat}=\sum_{l=0}^{L-1}\hat{h}_l\J_l\in\Complex^{M\times M}$ is the circulant matrix with first column $[\hat{\h}\tran \ \0_{M-L}\tran]\tran$, and
$\hat{\X}\in\Complex^{M\times K}$ is such that $\hat{\x}=\vvec(\hat{\X})$.
Given \eqref{eq:zns}-\eqref{eq:zss}, the structure of $\hat{\Hmat}$ and $\J_l$ 
imply
\begin{align}
\|\hat{\z}^{(n,*)}\|^2 & = \|\hat{\h}\|^2~\forall n   \label{eq:zns2} \\
\|\hat{\z}^{(*,l)}\|^2 & = \|\hat{\x}\|^2 = \|\hat{\X}\|_F^2~\forall l  \label{eq:zsl2}\\
\|\z^{(n,l)}\|^2 & = 1~\forall n,l                    \label{eq:znl2} .
\end{align}

With \eqref{eq:zss}-\eqref{eq:znl2}, PBiGAMP steps (R4)-(R6) reduce to
\begin{align}
\barnup  &= \nux\|\hat{\h}\|^2+ \frac{L}{MK}\nuh \|\hat{\x}\|^2 \\
\nup        &= \barnup + L\nux\nuh \\
\hat{\p}     &= \vvec(\hat{\Hmat}\hat{\X}) - \barnup \hat{\s} .
\end{align}
Furthermore, because $\hat{\Hmat}$ is circulant, its eigendecomposition takes the form 
\begin{align}
\hat{\Hmat} = \sqrt{M}\F_M\herm\Diag(\F_M^{1:L}\hat{\h})\F_M
\label{eq:circ}
\end{align}
after which the frequency-domain quantities
\begin{align}
\underline{\hat{\X}} 
&\triangleq  \F_M\hat{\X}  \label{eq:D_Xhat} \\
\underline{\hat{\h}} 
&\triangleq  \F_M^{1:L}\hat{\h} \label{eq:D_hhat}  
\end{align}
can be used to rewrite $\hat{\p}$ as
\begin{align}
\hat{\p} 
&= \vvec\big( \sqrt{M}\F_M\herm\Diag(\underline{\hat{\h}})\underline{\hat{\X}} \big)
        - \barnup \hat{\s} \label{eq:phat} .
\end{align}

Next we discuss PBiGAMP's nonlinear steps (R7)-(R8), which---according to (D1)---compute the posterior mean and variance of $\rvz_m$ given the likelihood function $p_{\rvy_m|\rvz_m}(y_m|z_m)$ from \eqref{eq:pyzm} and the prior $\rvz_m\sim \mathcal{CN}(\hat{p}_m,\nup)$.
Recall that the real and imaginary parts of $\mathcal{CN}(\hat{p}_m,\nup)$ are independent Gaussian with means $\hat{p}_m\re$ and $\hat{p}_m\im$, respectively, and variance $\nup/2$. 
Then, because the quantization $\mathcal{Q}(\cdot)$ is applied separately to real and imaginary components, we can separately compute the posterior means and variances for the real and imaginary components of $\rvz_m$.
Using $(g_{u-1},g_u]\subset\Real$ to denote the interval of $u_m\re$ quantized to $y_m\re$, the posterior mean and variance of the real part of $\rvz_m$ can be expressed as
\begin{align}
\hat{z}_m\re
&= \hat{p}_m\re + \frac{\nup}{2}\frac{D_m\re}{E_m\re} \label{eq:zhat} \\ 
\nu_m^{\text{z,re}}
&= \frac{\nup}{2} 
   + \frac{F_m\re}{E_m\re}\bigg(\frac{\nup}{2}\bigg)^2 
   - (\hat{z}_m\re-\hat{p}_m\re)^2 
  \label{eq:nuz}
\end{align}
where
\begin{align} 
D_m\re
&= \mathcal{N}\big(\hat{p}_m\re-g_{u-1}; 0,(\sigma^2_w+\nup)/2 \big) 
\nonumber\\&\quad 
   -\mathcal{N}\big(\hat{p}_m\re-g_{u}; 0,(\sigma^2_w+\nup)/2 \big) \\
E_m\re
&= \Phi\bigg(\frac{\hat{p}_m\re - g_{u-1}}{\sqrt{(\sigma^2_w+\nup)/2}}\bigg)   
- \Phi\bigg(\frac{\hat{p}_m\re - g_{u-1}}{\sqrt{(\sigma^2_w+\nup)/2}}\bigg) 
\label{eq:C} \\
F_m\re
&= \frac{\hat{p}_m\re- g_{u}}{(\sigma_w^2+\nup)/2}\mathcal{N}\big(\hat{p}_m\re-g_u; 0 ,(\sigma_w^2+\nup)/2 \big) 
\nonumber\\&\quad 
- \frac{\hat{p}_m\re- g_{u-1}}{(\sigma_w^2+\nup)/2}\mathcal{N}\big(\hat{p}_m\re-g_{u-1}; 0,  
         (\sigma_w^2+\nup)/2 \big)  \label{eq:D} .
\end{align}
Similarly, the posterior mean and variance of the imaginary part of $\rvz_m$ can be computed using the same procedure, but with $\hat{p}_m\im$ replacing $\hat{p}_m\re$.
Finally, for (R7)-(R8), the real and imaginary parts are combined as
\begin{align}
\hat{z}_m = \hat{z}_m\re + \text{j}\hat{z}_m\im, \quad
\nuz = \frac{1}{MK}\sum_{m=0}^{MK-1}\big(\nu^{\text{z,re}}_m + \nu^{\text{z,im}}_m \big).
\end{align}
Equations \eqref{eq:zhat}-\eqref{eq:D} can be derived following the procedures in \cite[Chapter 3.9]{Rasmussen:Book:06}; see \cite[Appendix A]{Wen:TSP:16} for further details.

Next we consider PBiGAMP steps (R11)-(R14). 
From \eqref{eq:zns}-\eqref{eq:zsl}, steps (R11) and (R13) become
\begin{align}
\nur &= \frac{1}{\nus\|\hat{\x}\|^2} \label{eq:rvar}\\
\nuq &= \frac{1}{\nus\|\hat{\h}\|^2}\label{eq:qvar}.  
\end{align}
For step (R12), we use \eqref{eq:zsl} and \eqref{eq:znl2} to write 
\begin{align}
\hat{r}_l 
&= \hat{h}_{l} + \nur\hat{\z}^{(*,l)\textsf{H}}\hat{\s} 
        - \nur\nus\nux\hat{h}_{l} \sum_{n=0}^{MK-1} \|\z^{(n,l)}\|^2  \\
&=  \hat{h}_{l}(1-MK\nur\nus\nux) + \nur \vvec(\J_l\hat{\X})\herm \vvec(\hat{\Smat}) \\
&=  \hat{h}_{l}(1-MK\nur\nus\nux) + \nur \sum_{k=1}^K (\J_l\hat{\x}_k)\herm \hat{\s}_k ,
\end{align} 
where $\hat{\Smat}\in\Complex^{M\times K}$ is a reshaping of $\hat{\s}$ and where
$\hat{\x}_k$ and $\hat{\s}_k$ are the $k$th columns of $\hat{\X}$ and $\hat{\Smat}$.
Thus $\rhat\defn[\hat{r}_0,\dots,\hat{r}_{L-1}]\tran$ takes the form
\begin{align}
\rhat 
&= \hat{\h}(1-MK\nur\nux\nus) 
   + \nur \sum_{k=1}^K \big[ \J_0\hat{\x}_k,\dots,\J_{L-1}\hat{\x}_k \big]\herm \hat{\s}_k 
\label{eq:rhat1} .
\end{align}
Since $\big[ \J_0\hat{\x}_k,\dots,\J_{L-1}\hat{\x}_k \big]$ are the first $L$ columns of the circulant matrix with first column $\hat{\x}_k$, \eqref{eq:circ} implies
\begin{align}
\big[ \J_0\hat{\x}_k,\dots,\J_{L-1}\hat{\x}_k \big]
&= \sqrt{M} \F_M\herm \Diag(\F_M\hat{\x}_k) \F_M^{1:L} 
\label{eq:Jx}.
\end{align}
Plugging \eqref{eq:Jx} into \eqref{eq:rhat1}, and defining $\underline{\hat{\x}}_k\defn \F_M\hat{\x}_k$ (i.e., the $k$th column of $\hat{\X}$) and $\underline{\hat{\s}}_k\defn \F_M\hat{\s}_k$, we get 
\begin{align}
\rhat 
&= \hat{\h}(1-MK\nur\nux\nus) 
   + \sqrt{M} \nur (\F_M^{1:L})\herm\sum_{k=1}^K\underline{\hat{\x}}_k^*\odot \underline{\s}_k 
\label{eq:rhat}	.
\end{align}
A similar derivation reduces PBiGAMP step (R14) to
\begin{align}
\qhat 
&=\hat{\x}(1-L\nuq\nuh\nus) + \sqrt{M}\nuq \vvec\big(
        \F_M\herm \Diag(\underline{\hat{\h}})\herm  \underline{\hat{\Smat}}\big) 
\label{eq:qhat} ,
\end{align}
where $\underline{\hat{\Smat}}\defn \F_M\hat{\Smat}$.

Next we consider PBiGAMP steps (R15)-(R16), which---according to (D2)---compute the posterior mean and variance of $\rvh_l$ given the GMM prior \eqref{eq:GMM} and the 
likelihood function $\mathcal{CN}(\hat{r}_l;h_l,\nur)$.
From \cite{Vila:TSP:13}, the posterior is 
\begin{align}
p_{\rvh_l|\rvr_l\!}(h_l\giv\hat{r}_l;\nur_l) 
&= \sum_{d=1}^D \bar{\lambda}_{l,d}\CN\Big(h_l; \frac{\nu_{l,d}\hat{r}_l}{\nu_{l,d}+\nur_l}, \frac{\nu_{l,d}\nur_l}{\nu_{l,d}+\nur_l}\Big) \qquad\\
\bar{\lambda}_{l,d}
&= \frac{\lambda_{l,d} \CN(\hat{r}_l; 0 , \nu_{l,d}+\nur_l)}
         {\sum_{d'=1}^D\lambda_{l,d'}\CN(\hat{r}_l; 0, \nu_{l,d'}+\nur_l)} 
         \label{eq:lambdabar},
\end{align}
which is also a GMM.
The corresponding mean and variance follow straightforwardly as
\begin{align}
\hat{h}_l 
&=\sum_{d=1}^D\bar{\lambda}_{l,d}\frac{\nu_{l,d}\hat{r}_l}{\nu_{l,d}+\nur_l}
\label{eq:hhat} \\
\nuh_l 
&=\sum_{d=1}^D\bar{\lambda}_{l,d} \left( \frac{\nu_{l,d}\nur_l}{\nu_{l,d}+\nur_l}+\Big|\frac{\nu_{l,d}\hat{r}_l}{\nu_{l,d}+\nur_l}\Big|^2 \right) -|\hat{h}_l|^2
\label{eq:hvar} .
\end{align}

Finally, we consider PBiGAMP steps (R17)-(R18), which---according to (D3)---compute the posterior mean and variance of $\rvx_n$ given the discrete symbol prior \eqref{eq:s_prior} and the likelihood function $\mathcal{CN}(\hat{q}_n;x_n,\nuq)$.
In this case, the posterior is
\begin{align}
\lefteqn{ p_{\rvx_n|\rvq_n\!}(x_n\giv\hat{q}_n;\nuq_n) 
= \sum_{j=1}^{2^A} \bar{\gamma}_{n,j}\delta(x_n-s^{(j)}) }\quad
\label{eq:s_post}\\
\bar{\gamma}_{n,j} 
&= \frac{\Pr\{\rvx_n\!=\!s^{(j)}\}\CN\big(s^{(j)};\hat{q}_n,\nuq_n\big)}
        {\sum_{j^\prime =1}^{2^A} \Pr\{\rvx_n\!=\!s^{(j')}\} \CN\big(s^{(j')};\hat{q}_n,\nuq_n\big) }, \label{eq:gammabar}
\end{align}
which is a discrete distribution with support on $\mathcal{S}$.
The posterior mean and variance follow as
\begin{align}
\hat{x}_{n} &= \sum_{j=1}^{2^A} \bar{\gamma}_{n,j} s^{(j)}
\label{eq:xhat}\\
\nux_n &= \sum_{j=1}^{2^A} \bar{\gamma}_{n,j} |s^{(j)}-\hat{x}_{n}|^2 
\label{eq:xvar} .
\end{align}
Note that $\{\bar{\gamma}_{n,j}\}_{j=1}^{2^A}$ is the posterior pmf on $\rvx_n$. 
It can be converted to posterior pmfs on the coded bits $\{\rvc_{n,a}\}_{a=1}^A$ via
\begin{align}
\lefteqn{ \Pr\{\rvc_{n,a}\!=\!1 \,|\, \hat{q}_n\} 
= \sum_{j=1...2^A|c_a^{(j)}=1} \Pr\{\rvcc_n\!=\!\cc^{(j)} \,|\, \hat{q}_n\} }\\
&= \sum_{\scriptsize \begin{array}{c}j=1...2^A\\c_a^{(j)}=1\end{array}} \sum_{j'=1}^{2^A} 
        \underbrace{ \Pr\{\rvcc_n\!=\!\cc^{(j)}|\rvx_n\!=\!s^{(j')}\} }_{\displaystyle \delta_{j-j'}} 
        \underbrace{ \Pr\{\rvx_n\!=\!s^{(j')} \,|\, \hat{q}_n\} }_{ \displaystyle \bar{\gamma}_{n,j'} }\\
&= \sum_{j=1...2^A | c_a^{(j)}=1} \bar{\gamma}_{n,j} 
\label{eq:c_post} .
\end{align}

The PBiGAMP-based soft equalization procedure is summarized in \tabref{UVSCpbigamp}
using $(M\times K)$-matricized versions of $\hat{\p}$, $\qhat$, and $\hat{\x}$ denoted by $\hat{\Pmat}$, $\hat{\Q}$, and $\hat{\X}$, respectively. 
Its complexity is dominated by the $4K+2$ DFT-matrix multiplies in steps 
(E1), (E2), (E5), (E10), (E12), and (E14), which consume a total of $O(MK\log M)$ operations per iteration, 
or $O(\log M)$ operations per symbol per iteration, when an FFT is used.
All other lines in \tabref{UVSCpbigamp} consume a total of $O(MK)$ operations per iteration, or $O(1)$ operations per symbol per iteration.

For notational simplicity, the table does not reflect the fact that the first $\Kp$ columns of $\hat{\X}$ are known pilots and the last $\Ng$ elements of the remaining columns in $\hat{\X}$ are known guards.
For those known elements, the mean and variance computations in (E17)-(E18) can be omitted.
Likewise, there is no need to compute the first $\Kp$ columns of $\underline{\hat{\X}}$ in (E1) or the first $\Kp$ columns of $\underline{\hat{\Q}}$ in (E14), reducing the number of required FFTs by $2\Kp$.

\putTable{UVSCpbigamp}{Soft Equalization via Scalar-Variance PBiGAMP}{\scriptsize
\begin{equation*}
\begin{array}{|r@{\,}c@{\,}l@{\,}r|}\hline

 \multicolumn{3}{|l}{\textrm{Definitions:}}&\\[-1mm]
 ~p_{\rvz_m|\rvp_{m}\!}\big(z\giv\hat{p};\nup\big) 
 &\triangleq& \frac{p_{\rvy_{m}|\rvz_{m}\!}(y_{m} \giv z) \, \CN(z;\hat{p},\nup)}{\int p_{\rvy_{m}|\rvz_{m}\!}(y_{m} \giv z') \, \CN(z';\hat{p},\nup) \dif z'}    &\text{(D1)}\\ 
 p_{\rvh_{l}|\rvr_{l}\!}(h\giv\hat{r};\nur) 
	&\triangleq& \frac{p_{\rvh_{l}\!}(h) \, \CN(\hat{r};h,\nur)}{\int p_{\rvh_{l}\!}(h') \, \CN(\hat{r};h',\nur) \dif h'}&\text{(D2)}\\
 p_{\rvx_{n}|\rvq_{n}\!}(x\giv\hat{q};\nuq) 
	&\triangleq& \frac{p_{\rvx_{n}\!}(x) \, \CN(\hat{q};x,\nuq)}{\int p_{\rvx_n\!}(x') \, \CN(\hat{q};x',\nuq)\dif x'}&\text{(D3)}\\[2mm]
 
 \multicolumn{4}{|l|}{\textrm{Initialization:}}\\[-2mm]
    \x_{0\textrm{G}} &=& [\0_{\Nd}\tran,\xG\tran]\tran &\\
    \hat{\X}[1] &=&  \multicolumn{2}{@{}l|}{
                     \big[\x_{\textrm{P},1},\dots,\x_{\textrm{P},\Kp},
                         \x_{0\textrm{G}},\dots,\x_{0\textrm{G}}\big],~\nux[1] = \frac{\Kd \Nd}{MK}}\\ 
    \hat{\h}[1] &=& \hat{\h}_{\textrm{init}},~ \nuh[1]=\nuh_\textrm{init},~ \hat{\Smat}[0]=\0_{M\times K} &\\[1mm]
 \multicolumn{4}{|l|}{\textrm{For $t=1,\dots T_{\max}$}}\\
     
   \underline{\hat{\X}}[t] &=& \F_M\hat{\X}[t]     &\text{(E1)} \\[1mm]   

   \underline{\hat{\h}}[t] &=& \F_M^{1:L}\hat{\h}[t]   &\text{(E2)} \\[1mm]  
   
   \barnup[t] &=& \nux[t]\big\|\hat{\h}[t]\big\|^2 + \frac{L}{MK}\nuh[t]\big\|\hat{\X}[t]\big\|_F^2&\text{(E3)} \\[1mm]                              					 	      
   \nup[t] &=& \barnup[t] + L\nuh[t] \nux[t]& \text{(E4)}\\[1mm]
                
   \hat{\Pmat}[t] &=& \sqrt{M}\F_M\herm\Diag(\underline{\hat{\h}}[t]) \underline{\hat{\X}}[t]  
                      - \barnup[t]\hat{\Smat}[t\!-\!1]  & \text{(E5)}\\[1mm] 	 
    				 
   \nuz[t] &=& \frac{1}{MK}\sum_{m=1}^{M-1}\sum_{k=1}^{K} \var\{\rvz_{mk}\giv\hat{p}_{mk}[t];\nup[t]\} & \text{(E6)}\\[1mm] 
   
   \forall m,k\!: \hat{z}_{mk}[t] &=& \Ex[\rvz_{mk}\giv\rvp_{mk}\!=\!\hat{p}_{mk}[t];\nup[t]] 						& \text{(E7)}\\[1mm] 
                
   \nus[t] &=& \big(1 - \nuz[t] / \nup[t]\big)/ \nup[t] & \text{(E8)}\\[1mm]
   
   \hat{\Smat}[t] &=&  \big(\hat{\Z}[t] - \hat{\Pmat}[t]\big) / \nup[t] & \text{(E9)}\\[1mm] 
   
   \underline{\hat{\Smat}}[t] &=& \F_M\hat{\Smat}[t]   & \text{(E10)}\\[1mm] 
	
   \nur[t] &=& \big(\nus[t]\big\|\hat{\X}[t]\big\|_F^2 \big)^{-1} & \text{(E11)}\\[1mm]   	
   \rhat[t] &=&\nur[t] \sqrt{M} (\F_M^{1:L})\herm 
        \big( \underline{\hat{\X}}[t]^* \odot \underline{\hat{\Smat}}[t] \big) \1_K
                &\\&&\mbox{} 
                + \big(1- MK\nur[t]\nux[t]\nus[t]\big) \hat{\h}[t] &\text{(E12)}\\[1mm]
                    
   \nuq [t] &=& \big(\nus[t]\big\|\hat{\h}[t]\big\|^2 \big)^{-1}   &\text{(E13)}\\[1mm]
   \hat{\Q}[t] &=& \sqrt{M}\nuq[t]
                \F_M\herm\Diag(\underline{\hat{\h}}[t])\herm \underline{\hat{\Smat}}[t] 
                &\\&&\mbox{} 
                + \big(1-L\nuq[t]\nuh[t]\nus[t]\big) \hat{\X}[t]  &\text{(E14)}\\[1mm]
    
    \nuh[t\!+\!1] &=& \frac{1}{L}\sum_{l=0}^{L-1}\var\{\rvh_{l}\giv \rvr_l=\hat{r}_{l}[t]; \nur[t]\} &   
   				     \text{(E15)}\\[0.5mm] 
    \forall l\!: \hat{h}_{l}[t\!+\!1]    
   				&=& \Ex[\rvh_l \giv \rvr_l \!=\!\hat{r}_{l}[t]; \nur[t]]& \text{(E16)}\\[0.5mm] 
    
    \nux[t\!+\!1] 
   				&=& \frac{1}{MK}\sum_{m=0}^{M-1}\sum_{k=1}^{K}\var\{\rvx_{mk}\giv\hat{q}_{mk}[t]; \nuq[t]\} & 
   				     \text{(E17)}\\[0.5mm] 
    \forall m,k\!: \hat{x}_{mk}[t\!+\!1] 
   				&=& \Ex[\rvx_{mk}\giv \rvq_{mk}\!=\!\hat{q}_{mk}[t]; \nuq[t]]& \text{(E18)}\\[0.5mm] 	
 
 \multicolumn{4}{|l|}{\textrm{end}}\\\hline	
\end{array}
\end{equation*}
}

\subsection{Turbo Equalization} \label{sec:turbo}

As described in \secref{BP}, we would like to compute (approximate) posterior marginal bit probabilities $\{p(b_i|\y)\}_{i=1}^{\Nb}$ using the SPA, which is the usual approach to turbo equalization \cite{Koetter:SPM:04}.
Because exact SPA is intractable for the soft-equalization subgraph in \figref{fg}, we use the PBiGAMP approximation, as described in \secref{pbigamp}, on that subgraph.
We now detail the remaining steps in the SPA, for completeness.

Roughly speaking, messages are passed on the factor graph in \figref{fg} from the left to the right and back again.
One such forward-backward pass will be referred to as a turbo iteration.
During a single turbo iteration, soft equalization using PBiGAMP is alternated with soft decoding using a standard decoder/interleaver.
The SPA dictates that ``extrinsic'' information is passed between nodes on the graph and hence between the subgraphs in \figref{fg}.
For a discrete random variable, the extrinsic message is a pmf formed by dividing the posterior pmf by the prior pmf.
Additional details are given below.

During each turbo iteration, extrinsic information on the coded bits $\rvc_{n,a}$ is passed from the soft decoder to PBiGAMP, where it is treated as prior information in \eqref{eq:s_prior} to determine the symbol priors $\gamma_{n,j}$.
PBiGAMP is then run to convergence, generating the symbol posteriors $\bar{\gamma}_{n,j}$.
The symbol posteriors are used in \eqref{eq:c_post} to determine the coded-bit posteriors, which are then converted to extrinsic form and passed to the soft decoder.
The soft decoder accepts this extrinsic information from PBiGAMP, treating it as a prior on the coded bits.
It then computes posteriors on the coded bits, converts them to extrinsic form, and passes them to PBiGAMP for the next turbo iteration.

\subsection{Learning the Channel Prior} \label{sec:EM}

The GMM prior \eqref{eq:GMM} requires specification of the weights and variances $\{\blam_l,\bnu_l\}_{l=0}^{L-1}$.
In the simple case where the coefficients are modeled as identically distributed, the set $\{\blam_l,\bnu_l\}_{l=0}^{L-1}$ reduces to the pair $\blam,\bnu$.
The ``EM-GM-AMP'' paper \cite{Vila:TSP:13} showed how this pair can be learned from the observations $\y$ using a combination of EM and AMP, and \cite{Parker:JSTSP:16} showed how EM can be combined with PBiGAMP in a similar manner.
In \secref{results}, we investigate the performance of this EM-GM-PBiGAMP method on the channels described in \secref{60Gchannel} using GMM order $D=2$.
More generally, one could partition the coefficients $\{\rvh_l\}_{l=0}^{L-1}$ into subsets and learn a different weight and variance for each subset, as discussed in \cite{Schniter:JSTSP:11}.
Typically, the EM update is performed in line (E16) once per PBiGAMP iteration, so that the computational burden of EM is very minor.

\subsection{Scaling the Channel Estimate} \label{sec:scale}

With few-bit ADCs, channel amplitude information is degraded due to quantization (and completely lost in the case of a one-bit ADC).
Thus, we find that channel-estimation performance can be improved by appropriately scaling the channel estimate.
To do this, we exploit the fact that
\begin{align}
\lefteqn{ \Ex[\|\uu\|^2\giv \h] 
= \tr\{\Ex[\uu\uu\herm \giv \h]\} } \\
&= \tr\{(\I_K\otimes\Hmat) \Ex[\x \x\herm](\I_K\otimes\Hmat)\herm\}+ MK\sigma_w^2 \\
&= \sigma_x^2 \tr\{\I_K\otimes\Hmat\Hmat\herm\}+ MK\sigma_w^2 \\
&= K \sigma_x^2 \tr\{\Hmat\Hmat\herm\}+ MK\sigma_w^2 \\
&= MK \sigma_x^2 \|\h\|^2+ MK\sigma_w^2 
\end{align}
due to the circulant nature of $\Hmat$, and so 
\begin{align}
\|\h\| = \sqrt{\dfrac{\Ex[\|\uu\|^2\giv\h]/(MK) - \sigma_w^2}{\sigma_x^2}}
\label{eq:hnorm}.
\end{align}
Assuming that the average received-signal power $\Ex[\|\uu\|^2 \giv \h]/(MK)$ can be measured\footnote{To measure the average received-signal power, it suffices to use an ADC with a relatively low sampling rate, which is inexpensive in both cost and power consumption.} prior to the ADC (as is typically done as part of automatic gain control), the true channel norm can be computed from \eqref{eq:hnorm} and the channel estimate $\hat{\h}$ can be scaled so that its norm matches the true one.
We note that a similar technique was used in \cite{Mo:TSP:18}.
With PBiGAMP, we scale the output of line~(E16) in this manner at each iteration.
 

\section{Benchmark Methods} \label{sec:bench}

We now describe two methods that will be used later for performance evaluation: PBiGAMP with Bussgang linearization, and pilot-aided channel estimation plus LMMSE decoding.

\subsection{PBiGAMP with Bussgang Linearization} \label{sec:bussgang}

The PBiGAMP method proposed in \secref{JCEDpbigamp} uses a non-Gaussian likelihood function $p_{\rvy_m|\rvz_m}$ that results directly from the quantization model \eqref{eq:quant_scalar}.
An alternative explored in the literature is the use of an AWGN approximation of $p_{\rvy_m|\rvz_m}$ based on a Bussgang linearization \cite{Mezghani:ISIT:12}.
This leads to a simplified approach that tends to perform well under mild quantization.
We briefly summarize the Bussgang approach below.\footnote{Our summary includes an explanation of why the effective noise $\tilde{\w}$ is uncorrelated with the signal $\x$, which is missing from \cite{Mezghani:ISIT:12}, as well as  
specializations relevant to \eqref{eq:vec_unquant}.}

The Bussgang linearization first writes the nonlinear quantization operation $\y=\mathcal{Q}(\uu)$ as
\begin{align}
\y
&= \Gy \uu + \e 
\label{eq:Bussgang},
\end{align}
where $\Gy$ is the LMMSE estimator of $\y$ from $\uu$, i.e.,
\begin{align}
\Gy = \Ex[\y\uu\herm] \Ex[\uu\uu\herm]^{-1} 
\label{eq:Fy} ,
\end{align}
and $\e\defn \y-\Gy\uu$ is the estimation error.
Due to the orthogonality principle, we know that $\Ex[\uu\e\herm]=\0$, i.e., the Bussgang error $\e$ is uncorrelated with the quantizer input $\uu$. 

Plugging the expression for $\uu$ from \eqref{eq:vec_unquant} into \eqref{eq:Bussgang}, we get
\begin{align}
\y
&= \Gy (\I_K \otimes \Hmat)\x
   + \underbrace{\Gy \w + \e}_{\displaystyle \defn \tilde{\w}}
\label{eq:AWGN} ,
\end{align}
where we can interpret $\Gy(\I_K \otimes \Hmat)$ as the effective channel and $\tilde{\w}$ as the effective noise.
Although non-Gaussian, $\tilde{\w}$ is approximately uncorrelated with the signal $\x$, in that
\begin{align}
\Ex[\x\tilde{\w}\herm]
&= \Ex[\x\w\herm]\Gy\herm + \Ex[\x\e\herm] \\
& = \Ex[\x\e\herm] 
\label{eq:uncorr1}\\
&= \Ex\big\{ \Ex[\x\e\herm|\uu]  \big\}
= \Ex\big\{ \Ex[\x|\uu] \e\herm \big\}
\label{eq:uncorr2}\\
&\approx \Ex[ \Gx\uu \e\herm ]
= \Gx \Ex[ \uu \e\herm ]
\label{eq:uncorr3}\\
&= \0 
\label{eq:uncorr4} ,
\end{align}
where 
\eqref{eq:uncorr1} follows from $\Ex[\x\w\herm]=\0$,
\eqref{eq:uncorr2} follows from the fact that $\e=\mathcal{Q}(\uu)-\Gy\uu$ is deterministic when conditioned on $\uu$, and
\eqref{eq:uncorr3} approximates $\Ex[\x|\uu]$ by the LMMSE estimate $\Gx\uu$ of $\x$ from $\uu$.
This approximation becomes exact when $\x$ and $\uu$ are jointly Gaussian.
Finally, equation \eqref{eq:uncorr4} follows from $\Ex[\uu\e\herm]=\0$.
 
Note that $\w$ and $\e$ are also uncorrelated, in that
\begin{align}
\Ex[\w\e\herm]
&= \Ex\big[ \Ex[\w\e\herm|\uu] \big] \\
&= \Ex\big[ \Ex[\w|\uu] \e\herm \big] 
\label{eq:uncorr5}\\
&= \Ex[ \Gw\uu \e\herm ]
= \Gw \Ex[ \uu \e\herm ]
\label{eq:uncorr6}\\
&= \0 
\label{eq:uncorr7} ,
\end{align}
where
\eqref{eq:uncorr5} results because $\e$ is deterministic conditioned on $\uu$,
\eqref{eq:uncorr6} results because $\w$ and $\uu$ are jointly Gaussian, with $\Gw$ denoting the LMMSE estimator of $\w$ from $\uu$,
and \eqref{eq:uncorr7} follows from $\Ex[\uu\e\herm]=\0$.
As a consequence of \eqref{eq:uncorr7}, the covariance of $\tilde{\w}$ reduces to 
\begin{align}
\Ex[\tilde{\w}\tilde{\w}\herm]
&= \sigma_w^2 \Gy\Gy\herm + \Ex[\e\e\herm] 
\label{eq:twcov} .
\end{align}

For uniform quantization with MMSE stepsize $\Delta_b$ \cite{Bucklew:TIT:80} (recall \eqref{eq:quant_scalar}), the LMMSE matrix $\Gy$ has a simple form.
To see this, we first define the quantization error 
\begin{align}
\q &\defn \y - \uu 
\label{eq:q} .
\end{align}
Note, from \eqref{eq:vec_unquant} and the fact that $\Hmat$ is circulant with first column $\h$, that $u_m=\sum_{l=0}^{M-1} h_{\langle m-l \rangle_M} x_{\lfloor m/M \rfloor M+l}$, where $\langle n \rangle_M$ denotes $n$-modulo-$M$.
Thus, if we treat the components of $\x$ as i.i.d., then the components of $\uu$ will be identically distributed. 
Consequently, the components of $\y=\mathcal{Q}(\uu)$ will be identically distributed, as will those of $\q$.
In this case, the results in \cite{Mezghani:ISIT:12} imply
\begin{align}
\Ex[\uu\q\herm] 
&= -\eta\Ex[\uu\uu\herm] = \Ex[\q\uu\herm]
\label{eq:tyq}\\
\Ex[\q\q\herm] 
&\approx \eta\Ex[\uu\uu\herm] 
        - (1-\eta)\eta~\textrm{Nondiag}(\Ex[\uu\uu\herm]) 
\label{eq:qq0} \\
&= \eta^2 \Ex[\uu\uu\herm] + (1-\eta)\eta \Diag(\diag(\Ex[\uu\uu\herm]))
\label{eq:qq} ,
\end{align}
where
\begin{align}
\eta 
&\defn \frac{\Ex[|q_m|^2]}{\Ex[|u_m|^2]}.
\end{align}
The approximation \eqref{eq:qq0} would be exact if $q_m$ and $y_{m'}$ were jointly Gaussian for all $m\neq m'$.
From \eqref{eq:Fy}, we now see that
\begin{align}
\Gy
&= \Ex[(\uu+\q)\uu\herm] \Ex[\uu\uu\herm]^{-1} \\
&= (1-\eta) \I
\label{eq:Fy2} ,
\end{align}
where \eqref{eq:Fy2} follows from \eqref{eq:tyq}.

We can now compute the effective noise covariance \eqref{eq:twcov}.
Noting from \eqref{eq:Bussgang}, \eqref{eq:q}, and \eqref{eq:Fy2} that 
\begin{align}
\e 
= \y - \Gy\uu 
= \uu + \q - (1-\eta)\uu
= \eta\uu + \q ,
\end{align}
we have
\begin{align}
\Ex[\e\e\herm]
&= \Ex[(\eta\uu+\q)(\eta\uu+\q)\herm] \\
&= \eta^2 \Ex[\uu\uu\herm] 
   + \eta \Ex[\uu\q\herm] 
   + \eta \Ex[\q\uu\herm] 
   + \Ex[\q\q\herm] \\
&= \Ex[\q\q\herm] - \eta^2 \Ex[\uu\uu\herm] 
\label{eq:ecov1}\\
&= (1-\eta)\eta \Diag(\diag(\Ex[\uu\uu\herm]))
\label{eq:ecov} ,
\end{align}
where 
\eqref{eq:ecov1} follows from \eqref{eq:tyq}
and \eqref{eq:ecov} follows from \eqref{eq:qq}.
Since 
\begin{align}
\Ex[|u_m|^2]
&= \Ex\big\{ [\I \otimes \Hmat]_{m,:} \x\x\herm [\I \otimes \Hmat]_{m,:}\herm \big\}
        + \sigma_w^2 \\
&= \sigma_x^2 \Ex[ \|\h\|^2 ] + \sigma_w^2 
\label{eq:ty2},
\end{align}
equations \eqref{eq:twcov}, \eqref{eq:Fy2}, \eqref{eq:ecov}, and \eqref{eq:ty2} imply 
\begin{align}
\lefteqn{ \Ex[\tilde{\w}\tilde{\w}\herm] }\nonumber\\
&= (1-\eta)\eta(\sigma_x^2\Ex\{\|\h\|^2\} + \sigma_w^2)\I 
        + (1-\eta)^2 \sigma_w^2 \I \\
&= \underbrace{ (1-\eta)( \eta \sigma_x^2 \Ex\{\|\h\|^2\}
        + \sigma_w^2 ) }_{\displaystyle \defn \sigtil} \I 
\label{eq:twcov2}.
\end{align}
Note that, in practice, $\Ex[|u_m|^2]$ can be estimated by measuring the input power to the ADC.

Finally, plugging \eqref{eq:Fy2} into \eqref{eq:AWGN}, we get
\begin{align}
\y
&= (1-\eta) (\I_K \otimes \Hmat)\x + \tilde{\w}
\label{eq:AWGN2} .
\end{align}
For the Bussgang approximation, we use \eqref{eq:AWGN2}, while treating the non-Gaussian effective noise $\tilde{\w}$ as if it was AWGN with variance $\sigtil$ from \eqref{eq:twcov2}. 

In going from standard to Bussgang-linearized PBiGAMP, changes manifest only in
lines~(R7)-(R8) of \tabref{pbigamp}.  
In either case, the complexity of lines~(R7)-(R8) is $O(MK)$ operations per frame, or $O(1)$ operations per symbol, recalling the discussion at the end of \secref{joint_pbigamp}. 
So, like PBiGAMP, the complexity of Bussgang-linearized PBiGAMP is $O(\log M)$ operations per symbol.

\subsection{Pilot-aided Channel Estimation and LMMSE Decoding} \label{sec:lmmse}

A computationally simpler benchmark is as follows.
First, using the standard correlation-based approach that leverages the perfect aperiodic autocorrelation property of Golay sequences described in \cite[Sec.~7.3.3.1]{Rappaport:Book:14}, we obtain $\hat{\Hmat}$. 
Next, treating the channel estimate as if it were perfect, we perform linear-MMSE (LMMSE) turbo decoding on the Bussgang-linearized model \eqref{eq:AWGN2}.
Details on the latter are provided below.

For each turbo iteration, we first convert the extrinsic information output by the coder into the data-symbol pmfs $\gamma_{n,j}$ via \eqref{eq:s_prior}, and then we convert these pmfs into the prior symbol mean and variance vectors $\bmu$ and $\vv$ via \eqref{eq:xhat}-\eqref{eq:xvar}.
At the very first turbo iteration, however, we set $\mu_n=0$ and $v_n=1$ for data indices $n$ (assuming unit-variance symbols) and $\mu_n=x_n$ and $v_n=0$ for the pilot/guard indices $n$.
Next, we compute the LMMSE symbol estimates $\hat{\x}$ and posterior symbol variance vector $\bnu^x$ as
\begin{align}
\hat{\x}
&= \bmu + \G(\y-\A\bmu) 
\label{eq:LMMSE_x}\\
\bnu^x 
&= \vv - \diag(\G\A\Diag(\vv)) ,
\end{align}
where
\begin{align}
\A &\defn (1-\eta) (\I_K \otimes \hat{\Hmat}) \\
\G &\defn \Diag(\vv)\A\herm\big(\A\Diag(\vv)\A\herm + \sigtil\I\big)^{-1} \label{eq:LMMSE_G}.
\end{align}
We then convert the posterior mean and variance $\hat{\x}$ and $\bnu^x$ to extrinsic quantities by solving for the $\hat{q}_n$ and $\nuq_n$ that yield
$1/\nux_n = 1/\nuq_n + 1/v_n$ 
and 
$\hat{x}_n/\nux_n = \hat{q}_n/\nuq + \mu_n/v_n$,
which is accomplished by
\begin{align}
\nuq_n &= \frac{v_n \nux_n}{v_n-\nux_n} \\
\hat{q}_n &= \frac{\hat{x}_nv_n-\mu_n\nux_n}{v_n-\nux_n} .
\end{align}
Finally we convert the extrinsic means and variances $\hat{q}_n$ and $\nuq_n$ into extrinsic coded-bit probabilities using \eqref{eq:gammabar} and \eqref{eq:c_post}, and pass them to the decoder.
The decoder treats them as coded-bit priors, computes coded-bit posteriors, and passes the extrinsic information back to the LMMSE equalizer to begin the next turbo iteration. 

As a result of the matrix inverse in \eqref{eq:LMMSE_G},
the LMMSE scheme \eqref{eq:LMMSE_x}-\eqref{eq:LMMSE_G} incurs a complexity of $O(KM^3)$ multiplies per block of $KM$ symbols, 
or $O(M^2)$ multiplies per symbol.
Compared to the $O(\log M)$ per-symbol per-iteration complexity of PBiGAMP, this is not favorable with regards to the scaling versus $M$.
However, if in \eqref{eq:LMMSE_G} we approximate the vector $\vv$ by its average value, then the per-symbol complexity could be reduced to $O(\log M)$, since $\widehat{\Hmat}$ is circulant and thus amenable to fast convolution.
In particular, this LMMSE approximation would use $4K+1$ FFTs per symbol block
(i.e.,
$1$ to compute the eigenvalues of $\widehat{\Hmat}$,
$2K$ for the multiplication by $\A$ in \eqref{eq:LMMSE_x}, and
$2K$ for the multiplication by $\G$ in \eqref{eq:LMMSE_x}).
Since PBiGAMP uses $4K+2$ FFTs, its per-iteration complexity would be only slightly higher.
Of course, PBiGAMP performs several iterations.  Still, we show in \secref{runtime} that the total computational complexity of PBiGAMP is only a bit higher than the fast LMMSE scheme, in part because it requires fewer turbo iterations on average.


\section{Numerical Results} \label{sec:results}

We now present numerical results comparing the proposed PBiGAMP method with the benchmarks discussed in \secref{bench}.
As a reference, we also consider the performance of PBiGAMP with perfect channel-state information (PCSI).
In this latter case, PBiGAMP reduces to GAMP.

\subsection{Setup}

\begin{figure}
	\begin{tikzpicture}[scale=1.06, transform shape]
	    \node (N_STF)[rectangle,draw, color= black, fill = green!40!white,text height=0.25cm] at (0,0) {$~~$STF$~~$}; 
        \node (N_CE)[rectangle,draw, right=0cm of N_STF,color= black, fill = red!60!white,text height=0.25cm] {$~\ $CEF$~\ $};
        \node (N_He)[rectangle,draw, right=0cm of N_CE,color= black, fill = orange!10!white,text height=0.25cm] {Header}; 
        \node (N_Data)[rectangle,draw, right=0cm of N_He,color= black, fill = blue!60!white,text height=0.25cm]    {$~~~~~~~~$Data$~~~~~~~~$};
        \node (N_TRN)[rectangle,draw, right=0cm of N_Data, dashed ,color= black, fill = black!30!white,text height=0.25cm]    {$~$TRN$~$}; 
        \draw (3.55,-0.7) node {(a)}; 
        
		\node (N_cdot)[rectangle,draw,color= white, text width=0.45cm, text height=0.25cm, dashed] at (-0.3,-2) {$\cdots $};
		\node (N_cdott)  [scale=1.5, align=center] at (-0.3,-2)  {$\cdots$};
		\node (N_G0)[rectangle,draw, right=0cm of N_cdot,color= black, fill=green!40!white, text width=0.45cm, text height=0.2cm, dashed] {$\minus  \g_a$};
		\node (N_G1)[rectangle,draw, right=0cm of N_G0,color= black, fill = red!60!white,  text width=0.45cm, text height=0.2cm] {$\minus  \g_b$};         
        \node (N_G2)[rectangle,draw, right=0cm of N_G1,color= black, fill = red!60!white,  text width=0.45cm, text height=0.2cm] {$\minus  \g_a$};
        \node (N_G3)[rectangle,draw, right=0cm of N_G2,color= black, fill = red!60!white,  text width=0.45cm, text height=0.2cm] {$\g_b$};
        \node (N_G4)[rectangle,draw, right=0cm of N_G3,color= black, fill = red!60!white,  text width=0.45cm, text height=0.2cm] {$\minus  \g_a$};
        \node (N_G5)[rectangle,draw, right=0cm of N_G4,color= black, fill = red!60!white,  text width=0.45cm, text height=0.2cm] {$\minus  \g_b$};
        \node (N_G6)[rectangle,draw, right=0cm of N_G5,color= black, fill = red!60!white,  text width=0.45cm, text height=0.2cm] {$\g_a$};
        \node (N_G7)[rectangle,draw, right=0cm of N_G6,color= black, fill = red!60!white,  text width=0.45cm, text height=0.2cm] {$\minus  \g_b$};
        \node (N_G8)[rectangle,draw, right=0cm of N_G7,color= black, fill = red!60!white,  text width=0.45cm, text height=0.2cm] {$\minus  \g_a$};
        \node (N_G9)[rectangle,draw, right=0cm of N_G8,color= black, fill = red!60!white,  text width=0.45cm, text height=0.2cm] {$\minus  \g_b$};
            \draw[decoration={brace,raise=5pt},decorate]
  					(-0.65,-1.8) -- node[above=5pt] {STF} (0.75,-1.8);
  	    \draw[decoration={brace,raise=5pt},decorate]
  					(0.75,-1.8) -- node[above=5pt] {CEF} (7.05,-1.8);
  	    \draw[decoration={brace,mirror,raise=5pt},decorate]
  					(0.75,-2.2) -- node[below=5pt] {$M$} (3.52,-2.2);
  	    \draw (3.55,-2.7) node {(b)};     
  	    \draw[decoration={brace,mirror,raise=5pt},decorate]
  					(3.57,-2.2) -- node[below=5pt] {$M$} (6.35,-2.2);
  	    
  	    \node (N_D1)[rectangle,draw, color= black, fill = blue!60!white, text width=2.0cm, text height=0.19cm,align = center] at (0.5,-4) 
  	    			{$\x_{\textrm{D},1}$};
  	    \node (N_G1)[rectangle,draw, right=0cm of N_D1, color= black, fill = blue!20!white, text width=0.45cm, text height=0.23cm]	{$\xG$};
  	    \node(N_D2)[rectangle,draw,right=0cm of N_G1,color= black,fill = blue!60!white,text width=2.0cm,text height=0.19cm,align=center]	
  	    			{$\x_{\textrm{D},2}$};
  	    \node (N_G2)[rectangle,draw, right=0cm of N_D2, color= black, fill = blue!20!white, text width=0.45cm, text height=0.23cm]	{$\xG$};
		\node (N_cdot1)  [right=0.3cm of N_G2, scale=1.5, align=center]  {$\cdots$};
		\draw[decoration={brace,raise=5pt},decorate]
  					(-0.63,-3.8) -- node[above=5pt] {$M$} (2.31,-3.8);
  		\draw[decoration={brace,mirror,raise=5pt},decorate]
  					(-0.63,-4.2) -- node[below=5pt] {$N_D$} (1.63,-4.2);
  		\draw[decoration={brace,mirror,raise=5pt},decorate]
  					(1.63,-4.2) -- node[below=5pt] {$N_G$} (2.31,-4.2);
  		\draw (3.55,-4.7) node {(c)};  
	\end{tikzpicture}
    \caption{(a) SC packet structure in the IEEE 802.11ad standard, including
                 the Short Training Field (STF), 
                 Channel Estimation Field (CEF), 
                 Header field, Data field, and 
                 optional Training (TRN) field for beamforming;
             (b) inner structure of the CEF, constructed from 
                 length-128 Golay complementary sequences $\{\g_a,\g_b\}$; and
             (c) inner structure of the Data block, composed of
                 data sequences $\{\x_{\text{D},1},\x_{\text{D},2}\}$ and guard intervals $\xG$.}
	\label{fig:packet_struc}
\end{figure}
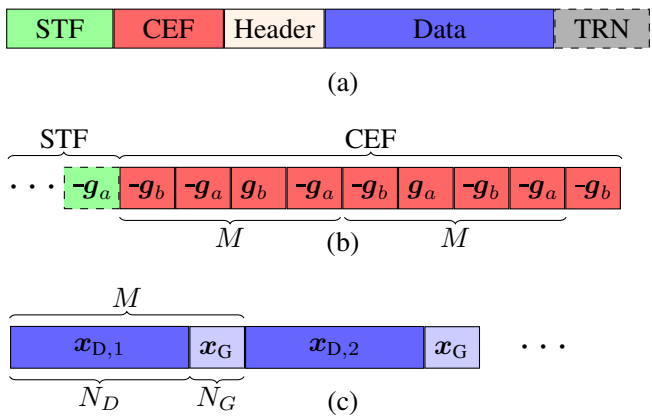

Unless otherwise noted, our numerical experiments are based on the following setup, which is compatible with the 802.11ad standard \cite{802.11ad}.
Recalling the SC block-transmission model from \secref{SC},
$\Nb=3584$ information bits were coded at rate $R=1/2$ by an irregular low-density parity-check (LDPC) code with average column weight $3$, as specified by \cite{802.11ad}. 
The $7168$ coded bits were then
Gray-mapped to $1792$ 16-QAM symbols (i.e., $A=4$). 
The data symbols were then partitioned into $\Kd = 4$ blocks of $\Nd = 448$ symbols, resulting in $\{\xD[k] \}_{k=1}^{4}$.
Each data-symbol sequence $\xD[k]$ was merged with an $\Ng=64$-length guard sequence $\xG$, resulting in a $M=512$-length data-guard sequence.
The set was then merged with $\Kp=2$ blocks of $M=512$ pilot symbols, 
as shown in Figs.~\ref{fig:block_struc} and \ref{fig:packet_struc}.

The 802.11ad standard specifies the use of Golay sequences \cite{Golay:IRETIT:61} for constructing both $\xP$ and $\xG$.
In particular, the pilot $\xP$ is constructed using the Golay complementary sequences $\{\g_a,\g_b\}$ as shown in \figref{packet_struc}(b), where both $\g_a$ and $\g_b$ have length $M/4 = 128$, and the guard $\xG$ is generated by an $\Ng=64$-length Golay sequence.
A correlation-based channel-estimation scheme that exploits the perfect aperiodic correlation property of Golay sequences is described in \cite[Sec.~7.3.3.1]{Rappaport:Book:14}.
We used that scheme for the benchmark described in \secref{lmmse}, as well as to initialize the proposed PBiGAMP approach. 

For the channel, we adopted the 60~GHz WLAN model described in \secref{60Gchannel}, 
whose Matlab implementation was obtained from \cite{Maslennikov:Tech:10}. 
We used the ``conference room'' scenario at baud rate 1.76~GHz with default parameter settings.
Interestingly, the delay spread of this channel exceeds the guard length ($\Ng=64$), implying some amount of inter-block interference (IBI).
However, the PDP in \figref{chan_realization}(b) suggests that the IBI power is relatively small. 

In the experiments below, one should remember that $E_b/N_o$ values correspond to post-beamforming SNRs, which include the gain of beamforming at both the transmitter and receiver.  
In multi-antenna systems, the pre-beamforming SNRs are much lower.

\subsection{BER and NMSE Performance with $\pi/2$-16-QAM} \label{sec:16QAM}

\begin{figure}[t]
    \centering
    \includegraphics[width=\figwid,clip]{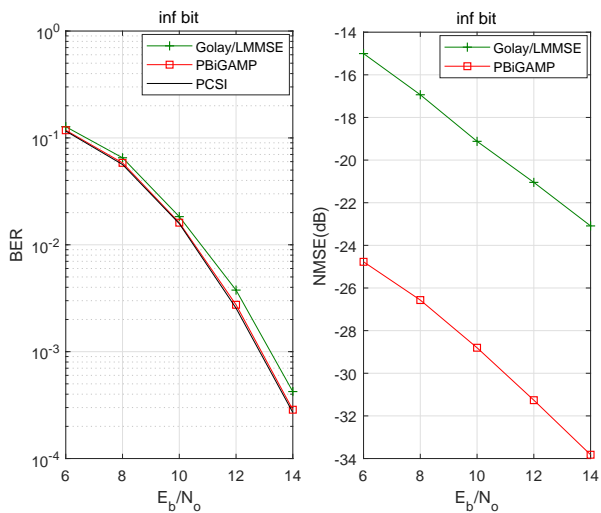}
    \caption{BER and channel NMSE versus $E_b/N_o$ in dB for 16-QAM with $\infty$-bit ADC under 60~GHz WLAN ``conference room'' channel.}
    \label{fig:16QAMinf}
\end{figure} 

\begin{figure}[t]
    \centering
    \includegraphics[width=\figwid,clip]{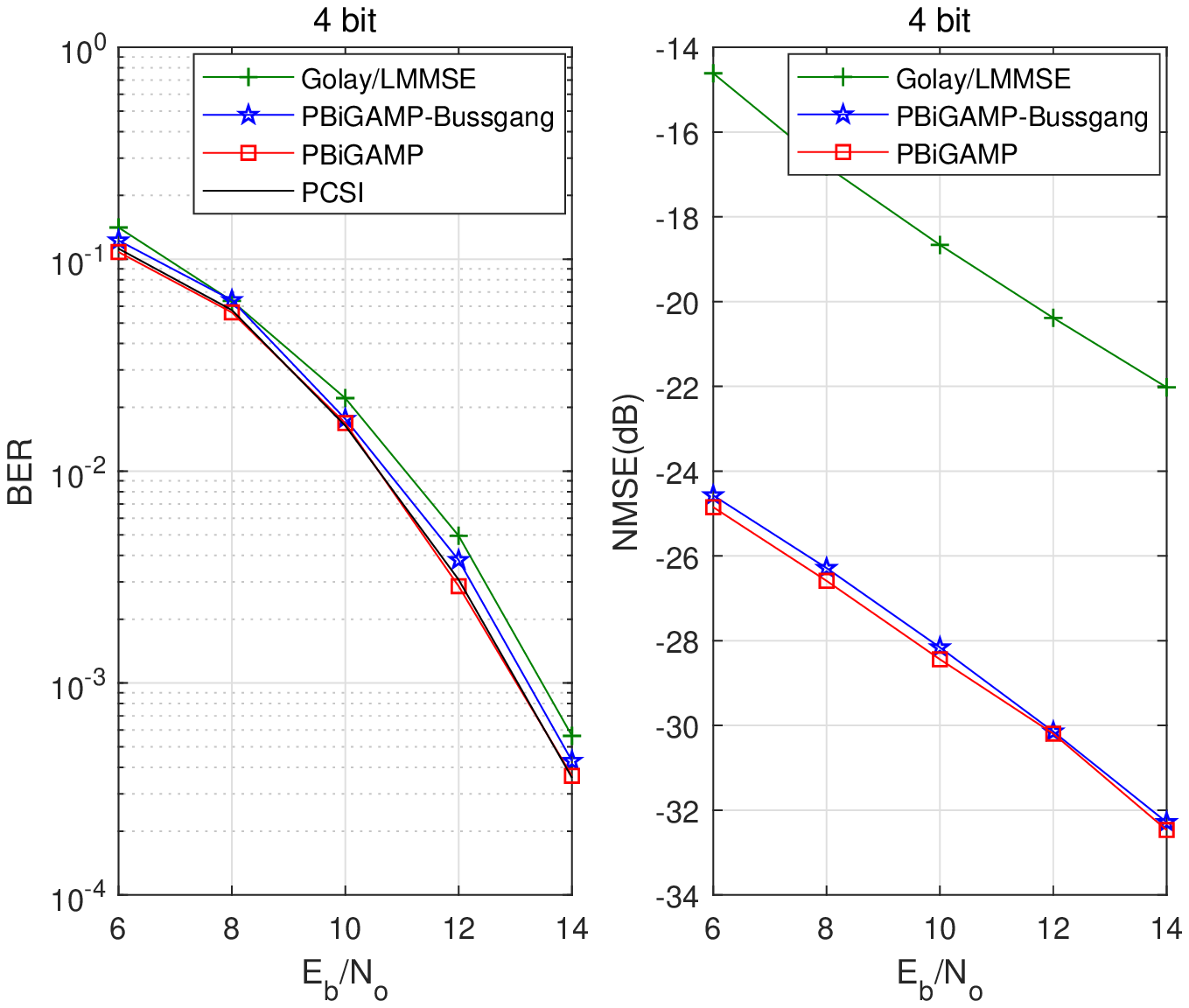}
    \caption{BER and channel NMSE versus $E_b/N_o$ in dB for 16-QAM with 4-bit ADC under 60~GHz WLAN ``conference room'' channel.}
    \label{fig:16QAM4}
\end{figure} 

\begin{figure}[t]
    \centering
    \includegraphics[width=\figwid,clip]{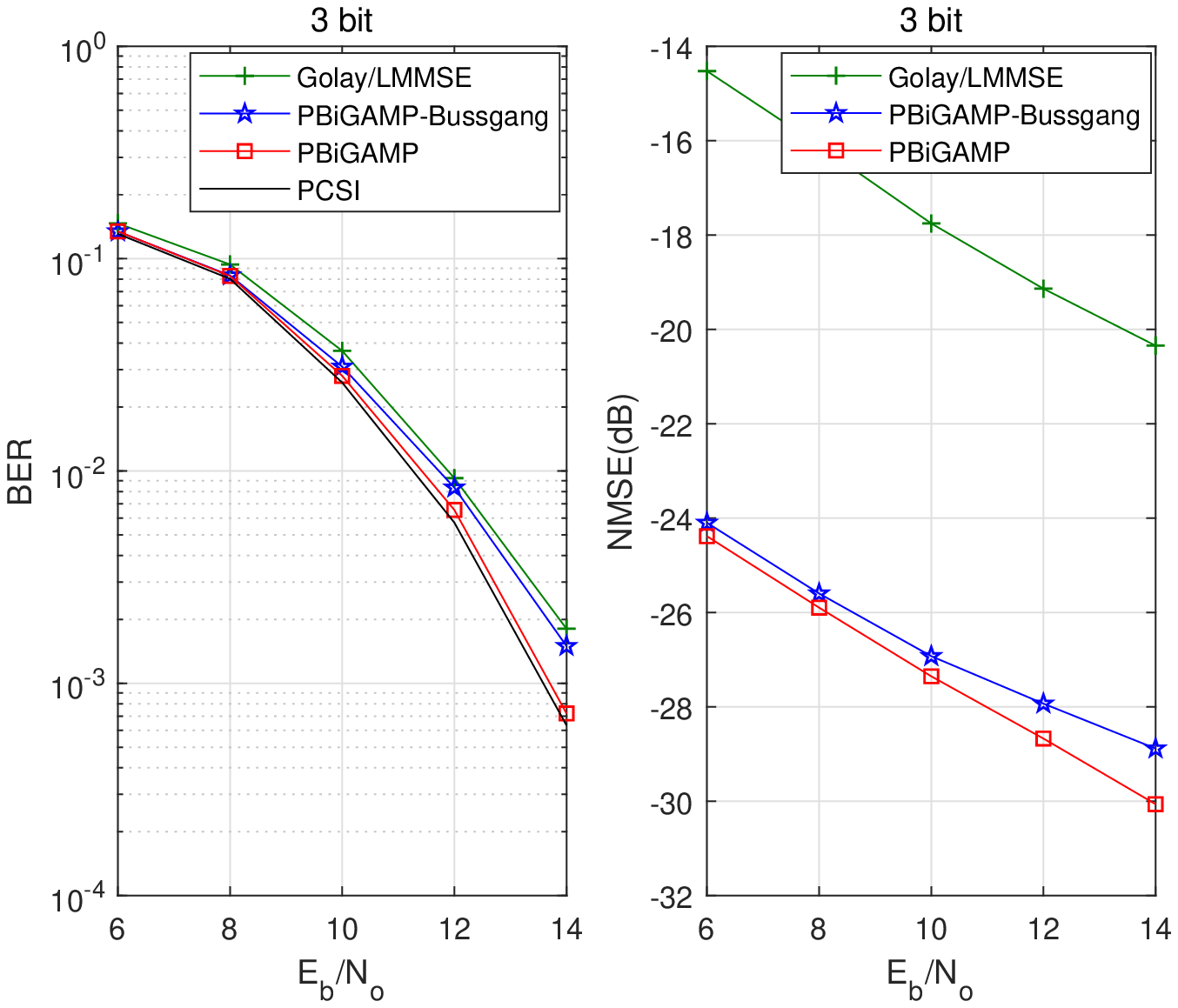}
    \caption{BER and channel NMSE versus $E_b/N_o$ in dB for 16-QAM with 3-bit ADC under 60~GHz WLAN ``conference room'' channel.}
    \label{fig:16QAM3}
\end{figure} 

\begin{figure}[t]
    \centering
    \includegraphics[width=\figwid,clip]{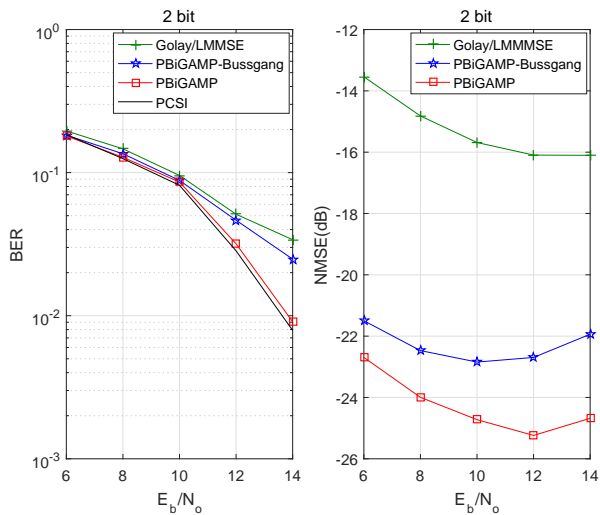}
    \caption{BER and channel NMSE versus $E_b/N_o$ in dB for 16-QAM with 2-bit ADC under 60~GHz WLAN ``conference room'' channel.}
    \label{fig:16QAM2}
\end{figure} 

Figures~\ref{fig:16QAMinf}-\ref{fig:16QAM2} show the bit error rate (BER) and the channel-estimation normalized MSE (NMSE) versus $E_b/N_o$ for ADCs with $\infty$-bit, 4-bit, 3-bit, or 2-bit precision.
With an $\infty$-bit ADC (i.e., no quantization), PBiGAMP achieves a BER that is nearly indistinguishable from the PCSI bound, while Golay/LMMSE is 0.4~dB worse in BER and 10~dB worse in NMSE.
With a 4-bit ADC the results are similar: PBiGAMP and PBiGAMP-Bussgang achieve BERs nearly indistinguishable from the PCSI bound (which has degraded 0.25~dB from the $\infty$-bit case), while Golay/LMMSE is 0.5~dB worse in BER and 10~dB worse in NMSE.
With a 3-bit ADC, PBiGAMP's BER is still nearly indistinguishable from the PCSI bound (which has degraded 0.8~dB from the $\infty$-bit case), while that of PBiGAMP-Bussgang is 0.7~dB worse and Golay/LMMSE is 0.9~dB worse in BER and 10~dB worse in NMSE.
With a 2-bit ADC, PBiGAMP's BER is still nearly indistinguishable from the PCSI bound (which has degraded 3.2~dB from the $\infty$-bit case), but the PBiGAMP-Bussgang and Golay/LMMSE BER traces show a large gap from the PCSI bound at high $E_b/N_o$.
The 2-bit NMSE traces are non-monotonic as a result of the ``stochastic resonance'' phenomenon \cite{Mo:ASIL:14,Mo:TSP:18}, referring to the phenomemon where noise improves the performance of a nonlinear system \cite{McDonnell:Book:08}.

\subsection{BER and NMSE Performance with $\pi/2$-BPSK} \label{sec:BPSK}

In our experiments with 1-bit ADC, we found that none of the schemes under test were able to reliably decode the 16-QAM transmission described in \secref{16QAM}.
We now show that 1-bit reception is feasible for $\pi/2$-BPSK transmissions, which is a mandatory mode of the 802.11ad standard \cite{802.11ad}.
For this, we coded $\Nb=896$ information bits as before (i.e., at rate $R=1/2$ using an irregular LDPC code with average column weight $3$). 
The $1792$ coded bits were then randomly interleaved and Gray-mapped to $\Nd=1792$ symbols using $\frac{\pi}{2}$-BPSK (which rotates a standard BPSK transmission by $\pi/2$ radians each baud interval for improved PAPR). 
All other settings were the same as described earlier.

\begin{figure}[t]
    \centering
    \includegraphics[width=\figwid,clip]{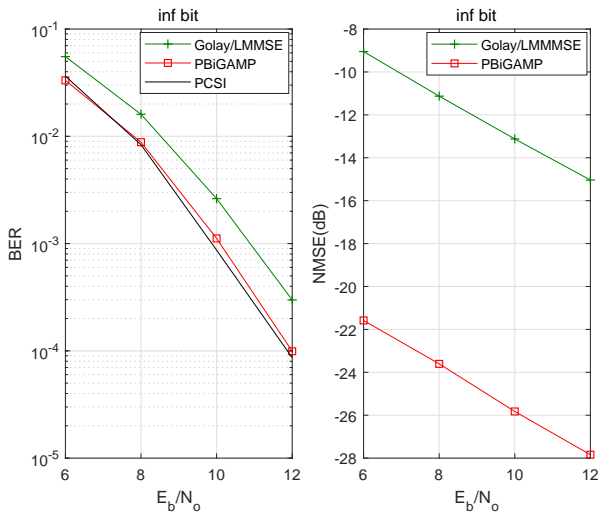}
    \caption{BER and channel NMSE versus $E_b/N_o$ in dB for $\pi/2$-BPSK with $\infty$-bit ADC under 60~GHz WLAN ``conference room'' channel.}
    \label{fig:BPSKinf}
\end{figure}  

\begin{figure}[t]
    \centering
    \includegraphics[width=\figwid,clip]{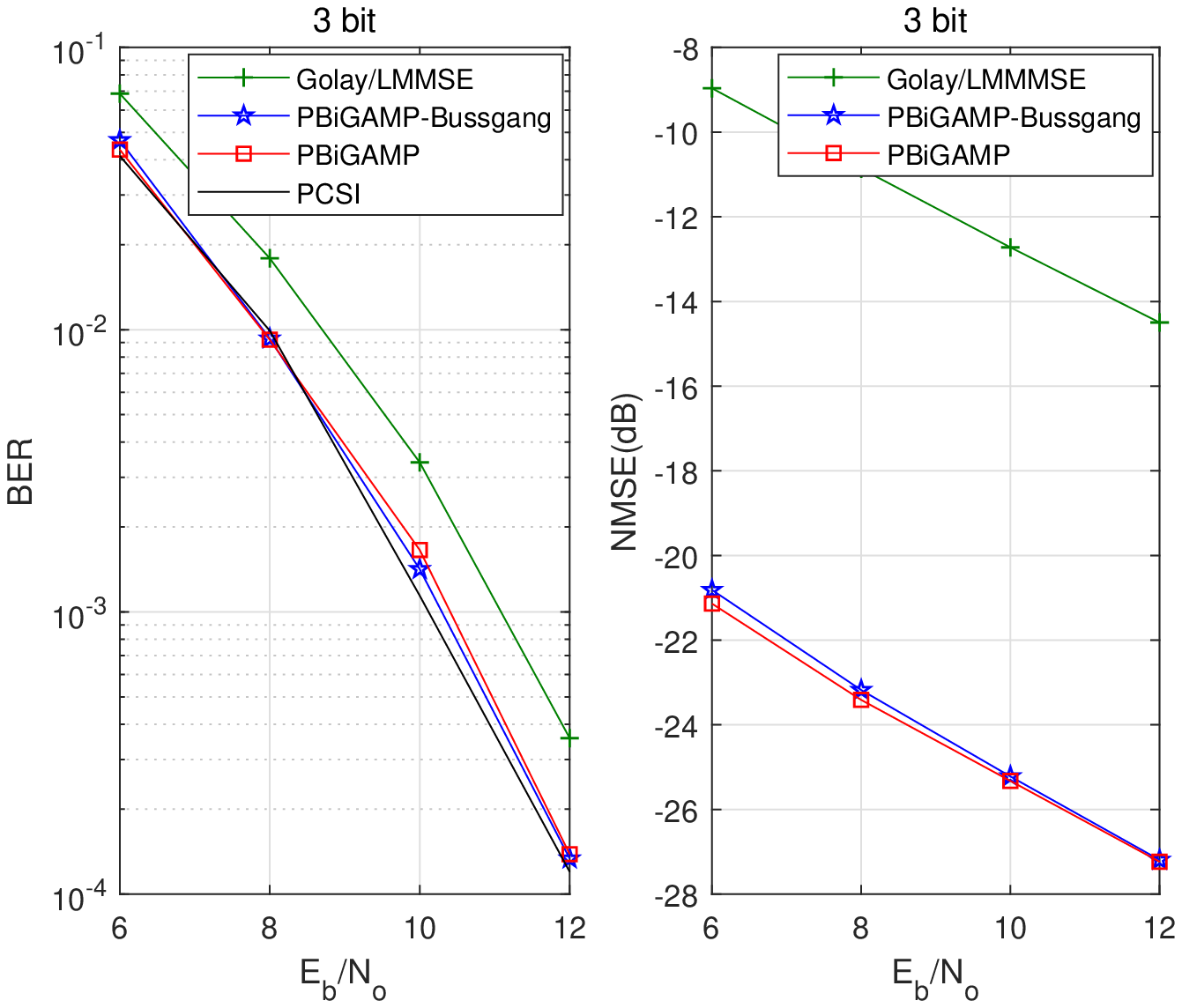}
    \caption{BER and channel NMSE versus $E_b/N_o$ in dB for $\pi/2$-BPSK with 3-bit ADC under 60~GHz WLAN ``conference room'' channel.}
    \label{fig:BPSK3}
\end{figure} 

\begin{figure}[t]
    \centering
    \includegraphics[width=\figwid,clip]{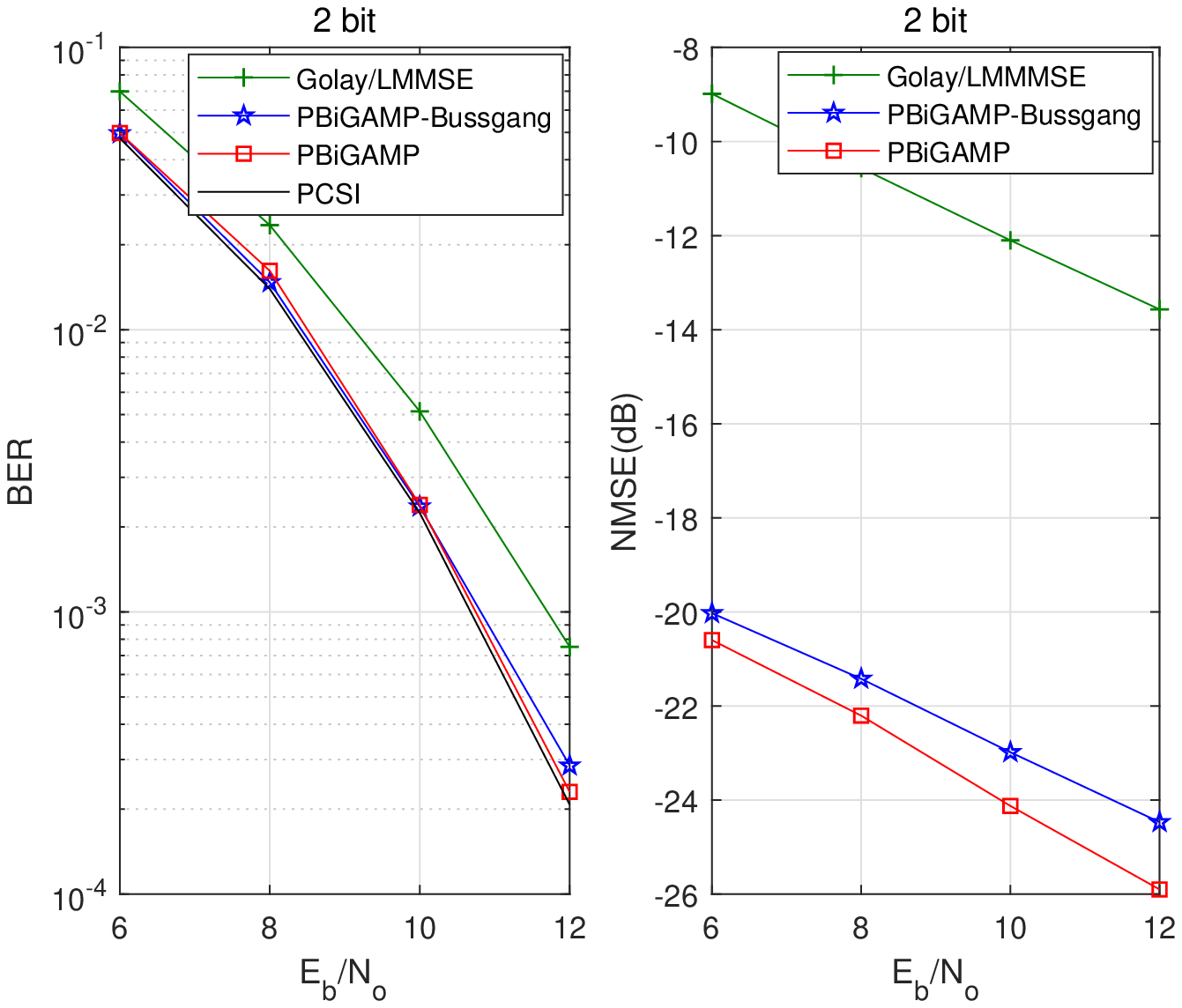}
    \caption{BER and channel NMSE versus $E_b/N_o$ in dB for $\pi/2$-BPSK with 2-bit ADC under 60~GHz WLAN ``conference room'' channel.}
    \label{fig:BPSK2}
\end{figure} 

\begin{figure}[t]
    \centering
    \includegraphics[width=\figwid,clip]{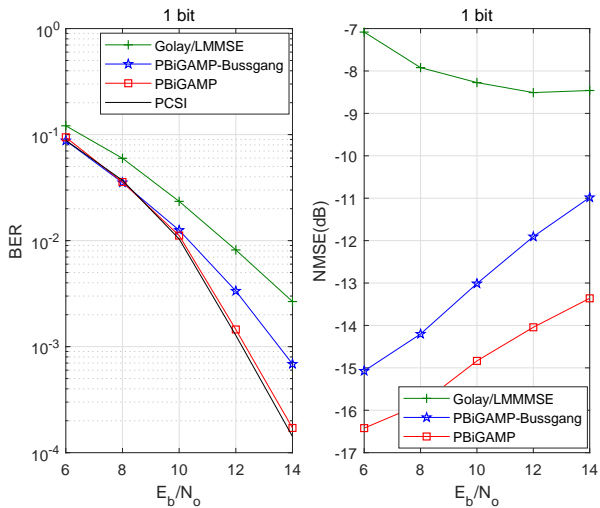}
    \caption{BER and channel NMSE versus $E_b/N_o$ in dB for $\pi/2$-BPSK with 1-bit ADC under 60~GHz WLAN ``conference room'' channel.}
    \label{fig:BPSK1}
\end{figure} 

Figures~\ref{fig:BPSKinf}-\ref{fig:BPSK1} show the bit error rate (BER) and the channel-estimation normalized MSE (NMSE) versus $E_b/N_o$ for ADCs with $\infty$-bit, 3-bit, 2-bit, and 1-bit precision, respectively.
With an $\infty$-bit ADC (i.e., no quantization), PBiGAMP achieves a BER that is nearly indistinguishable from the PCSI bound, while Golay/LMMSE is 0.9~dB worse in BER and 13~dB worse in NMSE.
With a 3-bit ADC the results are similar: PBiGAMP and PBiGAMP-Bussgang achieve BERs nearly indistinguishable from the PCSI bound (which has degraded 0.3~dB from the $\infty$-bit case), while Golay/LMMSE is 0.9~dB worse in BER and 13~dB worse in NMSE.
With a 2-bit ADC, the BERs of PBiGAMP and PBiGAMP-Bussgang are nearly indistinguishable from the PCSI bound (which has degraded 0.6~dB from the $\infty$-bit case), while Golay/LMMSE is 1~dB worse in BER and 13~dB worse in NMSE.
With a 1-bit ADC, PBiGAMP's BER is still nearly indistinguishable from the PCSI bound (which has degraded 2.2~dB from the $\infty$-bit case), but the PBiGAMP-Bussgang and Golay/LMMSE BER traces show a large gap from the PCSI bound at high $E_b/N_o$.
The 1-bit NMSE traces are non-monotonic as a result of the ``stochastic resonance'' phenomenon \cite{Mezghani:ISIT:12}.

\subsection{BER versus Runtime with 16-QAM} \label{sec:runtime}

To assess the computational complexity of PBiGAMP relative to the benchmark methods, we now present the results of runtime experiments in Matlab on a 3.3~GHz CPU.\footnote{The runtimes would be much faster in an ASIC or FPGA implementation.}
The algorithms under test were PBiGAMP, Bussgang-linearized PBiGAMP, the exact Golay/LMMSE scheme \eqref{eq:LMMSE_x}-\eqref{eq:LMMSE_G}, and the fast approximate Golay/LMMSE scheme described at the end of \secref{lmmse}.
PBiGAMP was terminated at the smallest iteration $t\geq 7$ at which $\sum_{m,k} |\hat{x}_{mk}[t\!+\!1]-\hat{x}_{mk}[t]|^2 < 0.01 \sum_{m,k} |\hat{x}_{mk}[t\!+\!1]|^2$.

\begin{figure}[t]
	\centering
        \newcommand{\sz}{0.55}
        \psfrag{PBiGAMP}[l][l][\sz]{\sf PBiGAMP}
        \psfrag{PBiGAMP-Busggang}[l][l][\sz]{\sf PBiGAMP-Bussgang}
        \psfrag{Golay/LMMSE-inverse}[l][l][\sz]{\sf Golay/LMMSE}
        \psfrag{Golay/LMMSE}[l][l][\sz]{\sf Golay/LMMSE-Fast}
	\includegraphics[width=\figwid,clip]{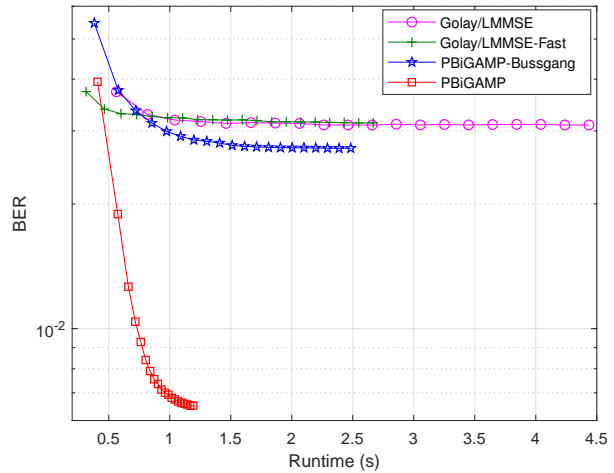}
	\caption{BER versus average runtime for several algorithms with 16-QAM modulation and 2-bit quantization at $E_b/N_o=$~14~dB.}
	\label{fig:runtime_2bit}
\end{figure}

\begin{figure}[t]
	\centering
        \newcommand{\sz}{0.55}
        \psfrag{PBiGAMP}[l][l][\sz]{\sf PBiGAMP}
        \psfrag{PBiGAMP-Busggang}[l][l][\sz]{\sf PBiGAMP-Bussgang}
        \psfrag{Golay/LMMSE-inverse}[l][l][\sz]{\sf Golay/LMMSE}
        \psfrag{Golay/LMMSE}[l][l][\sz]{\sf Golay/LMMSE-Fast}
	\includegraphics[width=\figwid,clip]{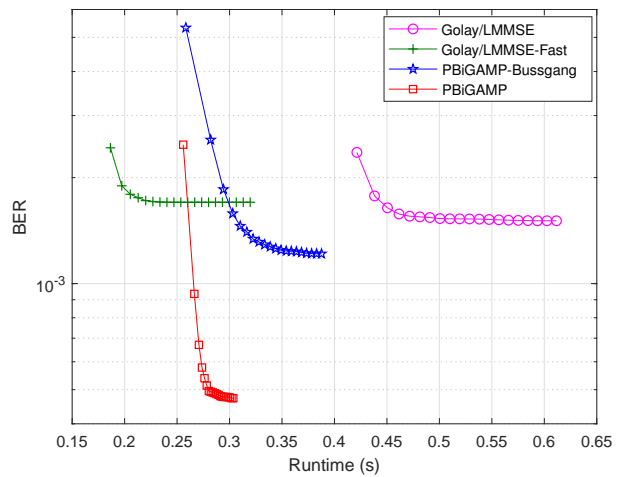}
	\caption{BER versus average runtime for several algorithms with 16-QAM modulation and 3-bit quantization at $E_b/N_o=$~14~dB.}
	\label{fig:runtime_3bit}
\end{figure}

Figures~\ref{fig:runtime_2bit} and \ref{fig:runtime_3bit} plot BER versus average runtime for 16-QAM modulation and $E_b/N_o=$~14~dB at 2-bit and 3-bit quantization, respectively.
The markers in each trace show the average BER and the average (cumulative) runtime at the end of each turbo iteration, indexed from 1 through 20.
For each Monte-Carlo trial, a parity check was used to determine whether the BER was zero at the beginning of each turbo iteration and, if so, the equalization and decoding operations in that iteration were skipped.
Thus, the \emph{average} runtime contribution of the $i$th turbo iteration decrease with the iteration index $i$, because it is more likely that the BER equals zero in later turbo iterations. 

\Figref{runtime_2bit} shows that, with 2-bit quantization, the fastest output comes from Golay/LMMSE-Fast after a single turbo iteration.
However, the corresponding BER is relatively poor.
At 2 turbo iterations, PBiGAMP yields a much lower BER than all other schemes, while consuming the same runtime as only 3 turbo iterations of Golay/LMMSE-Fast.
And PBiGAMP yields even lower BERs after $>$~2 turbo iterations.
Overall, \figref{runtime_2bit} shows that PBiGAMP's accuracy-complexity tradeoff is vastly superior to those of the other methods.

\Figref{runtime_3bit} shows similar behavior with 3-bit quantization.
As before, Golay/LMMSE-Fast achieves the fastest decoding, 
but its BER is relatively poor.
After only 2 turbo iterations, the BER of PBiGAMP surpasses the BERs achieved by all other methods.
And the time it takes for PBiGAMP to complete 2 turbo iterations is only about 40\% more than the time it takes for Golay/LMMSE-Fast to complete 2 turbo iterations.
So, PBiGAMP gives a significant improvement in BER for a modest increase in complexity.

Several other observations can be made from Figs.~\ref{fig:runtime_2bit}-\ref{fig:runtime_3bit}.
First the fast/approximate LMMSE scheme is much faster than the exact LMMSE scheme, although it yields slightly worse BER. 
Both behaviors are expected.
Second, lower BER translates to faster average runtime per turbo iteration, because fewer turbo iterations need to be performed.
So, more accurate equalization leads to improvements in runtime.

\subsection{Robustness to Noise-Variance Mismatch} \label{sec:robustness}

Recall that all methods under test take the noise variance $\sigma_w^2$ as an imput.
We now examine robustness to mismatch between the assumed and true values of $\sigma_w^2$.

\begin{figure}[t]
	\centering
        \psfrag{NV (dB)}[t][t][0.6]{\sf $\sigma_w^2$ mismatch (dB)}
        \newcommand{\sz}{0.5}
        \psfrag{PBiGAMP}[l][l][\sz]{\sf PBiGAMP}
        \psfrag{PBiGAMP-Busggang}[l][l][\sz]{\sf PBiGAMP-Bussgang}
        \psfrag{Golay/LMMSE}[l][l][\sz]{\sf Golay/LMMSE}
        \psfrag{PCSI}[l][l][\sz]{\sf PCSI}
	\includegraphics[width=\figwid,clip]{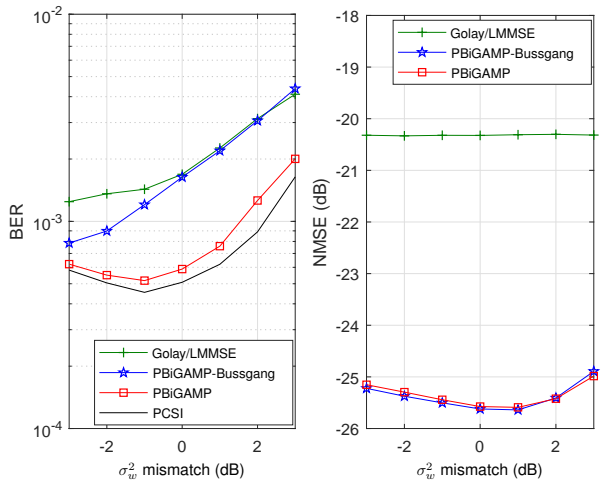}
	\caption{BER and channel NMSE versus noise-variance mismatch in dB for 16-QAM with 3-bit quantization under the 60~GHz WLAN ``conference room'' channel at $E_b/N_o=$~14~dB.}
	\label{fig:mismatch}
\end{figure}

\Figref{mismatch} shows the BER and channel-estimation NMSE versus $\sigma_w^2$-mismatch in dB for 16-QAM with 3-bit ADC quantization at $E_b/N_o=$~14~dB.
The figure shows that, as the assumed value of $\sigma_w^2$ grows larger than the true $\sigma_w^2$ (i.e., the mismatch in dB grows positive), the BERs of all methods degrade at a similar rate.
However, as the assumed value of $\sigma_w^2$ grows smaller than the true $\sigma_w^2$ (i.e., the mismatch in dB grows negative), the BERs of all methods slightly improve before finally degrading.
\Figref{mismatch} also shows that PBiGAMP's channel estimation NMSE slightly degrades in the presence of noise-variance mismatch, while that of the Golay/LMMSE scheme remains relatively constant (but far worse than the value achieved by PBiGAMP).

Importantly, the BER of PBiGAMP closely tracks that of the perfect-CSI benchmark over the entire range of mismatch.
This is the best possible outcome among schemes that take the noise variance $\sigma_w^2$ as an input parameter.
Of course, it would be better to learn $\sigma_w^2$ from $\y$ rather than trust the supplied value of $\sigma_w^2$.
As discussed in footnote~\ref{foot:noise}, while extending PBiGAMP to learn $\sigma_w^2$ should not be difficult, we leave it for future work.

\section{Conclusions} \label{sec:conc}

In this paper we proposed a fast and near-optimal approach to joint channel-estimation, equalization, and decoding of coded SC transmissions over frequency-selective channels with few-bit ADCs.
Our approach leverages the PBiGAMP algorithm to reduce the implementation complexity of joint channel estimation and symbol decoding to that of a few FFTs per iteration.
Furthermore, it learns and exploits sparsity in the channel impulse response.
Our work is motivated by millimeter-wave systems with bandwidths on the order of Gsamples/sec, where few-bit ADCs, SC transmissions, and fast processing all lead to significant reductions in power consumption and implementation cost.
We demonstrated our approach using signals and channels generated according to the IEEE 802.11ad wireless LAN standard, in the case that the receiver uses analog beamforming and a single ADC.
Our experiments showed that the proposed approach yields BER almost indistinguishable from the known-channel oracle for ADCs with as few as 2-bit precision when recovering coded 16-QAM transmissions, and for ADCs with as few as 1-bit precision when recovering coded BPSK transmissions.
Although it should be possible to recover coded QPSK transmissions with 1-bit ADCs, none of the schemes considered in this paper were able to do reliably with the 802.11ad codes and 802.11ad channels, and thus further work in this direction is warranted.
As future work, it would also be interesting to extend our method to learn the noise variance $\sigma_w^2$ and to work with multiple few-bit ADCs, as in digital or hybrid beamforming systems.


\bibliographystyle{IEEEtran}
\bibliography{macros_abbrev,comm,stc,multicarrier,sparse,books,misc,machine}
\end{document}